\documentclass[aps,reprint,amsmath, amssymb,groupedaddress,floatfix]{revtex4-1}
\usepackage{natbib}
\usepackage{graphicx}
\usepackage{bm}
\usepackage{color,soul}

\begin{document}
\preprint{APS/123-QED}
\title{Universality in the Electronic Structure of 3d Transition Metal Oxides}
\author{Priyadarshini Parida}
\author{Ravi Kashikar}
\author{Ajit Jena}
\author{B. R. K. Nanda}
\email{nandab@iitm.ac.in}
\affiliation{Condensed Matter Theory and Computational Lab, Department of Physics, \\ Indian Institute of Technology Madras, Chennai - 36, India}
\date{\today}

\begin{abstract}

Electronic structure of strongly correlated transition metal oxides (TMOs) is a complex phenomenon due to  competing interaction among the charge, spin, orbital and lattice degrees of freedom. Often individual compounds are examined to explain certain properties associated with these compounds or in rare cases few members of a family are investigated to define a particular trend exhibited by that family. Here, with the objective of generalization, we have investigated the electronic structure of three families of compounds, namely, highly symmetric cubic mono-oxides, symmetry-lowered spinels and asymmetric olivine phosphates,  through density functional calculations. From the results we have developed empirical hypotheses involving electron hopping, electron-lattice coupling, Hund's rule coupling, strong correlation and d-band filling. These hypotheses, classified through the point group symmetry of the transition metal - oxygen complexes, can be useful to understand and predict the electronic and magnetic structure of 3d TMOs.  
\end{abstract}

\maketitle

\section{Introduction}
\label{sec:intro}
Ground state electronic structure is an intriguing phenomenon in 3d transition metal oxides (TMOs). Here ten-fold degeneracy of 3d electrons and six-fold degeneracy of O-p electrons suffer  massive symmetry breakdown mediated by crystal field, orbital dependent covalent hybridization, Hund's coupling, on-site Coulomb repulsion, and valence electron count (VEC). In the case of partially filled d-shells, Hund's coupling breaks spin-symmetry and the ten-fold degeneracy gives rise to two five-fold degenerate states. Due to spatial inhomogeneity in the charge distribution of the d-orbitals, the crystal field exerted on them by the anions breaks the five-fold degeneracy further. For example, if the M-O complex is an octahedron as in the case of monoxides \cite{mattheiss} and perovskites \cite{sashi}, then the crystal field effect removes the five-fold degeneracy and splits the d-states into three-fold $t_{2g}$ states lying lower in energy and two-fold e$_g$ states lying higher in energy. However, the situation is reverse in the case of tetrahedral geometry (NiFe$_2$O$_4$ \cite{meinert}, Fe$_3$O$_4$ \cite{noh}, YCrO$_4$ \cite{tsirlin}, Sr$_3$Cr$_2$O$_8$ \cite{radtke}) where the doubly degenerate $e$ states are lower in energy than the triply degenerate $t_2$ states. Some of the systems like the spinel compounds (e.g. NiFe$_2$O$_4$ \cite{meinert} and Fe$_3$O$_4$ \cite{noh}) consist of both tetrahedra and octahedra where the electronic structure is governed by four sets of degenerate states. Further, depending on the occupancy of these states, the strong correlation effect can further separate these degenerate states into lower and upper Hubbard bands.  

The importance of crystal field effect and strong correlation effect in TMOs can be realized from the series of materials SrVO$_3$, CaVO$_3$, LaTiO$_3$, YTiO$_3$, having $d_1$ occupancy of the $t_{2g}$ shell, SrVO$_3$ and CaVO$_3$ are correlated metals whereas LaTiO$_3$ and YTiO$_3$ are reported to be Mott insulators. This consequence is attributed to the further breakdown of the octahedral field split  caused by orthorhombic distortion which increases as one moves along the series \cite{pavarini, pavarini1, george, inoue}. In La$_{1-x}$Ca$_x$MnO$_3$ \cite{pickett} and La$_{1-x}$Sr$_x$MnO$_3$ \cite{piskunov, pavone, paraskevopoulos, fang, chen3}, the structural distortion (cubic to orthorhombic) and d-occupancy varies with doping  concentration. As a consequence, variation of $x$ stabilizes different electronic and magnetic phases which include type A antiferromagnetic (A-AFM) and insulating phase at $x=0$, ferromagnetic (FM) and metallic phase at $x = 0.33$ and type G-AFM insulating phase at $x=1$. Furthermore, in the insulating phase, depending on $x$ three different types of charge ordering can be observed. 

So far most of electronic structure calculations using mean-field and statistical methods (e.g. density functional theory (DFT) and quantum Monte Carlo based solvable model tight-binding Hamiltonians) are carried out on individual compounds or few members of a given family to identify certain trend or certain effects that explain the experimental finding. While there are many studies on individual compounds to cite, very few literatures have investigated a whole family. Some of the  examples are Hartree-Fock study on RMO$_3$ (R = rare earth atom) and M = Ti, V, Mn, Ni) \cite{mizokawa}, DFT study on double-perovskites (Sr$_2$MMoO$_6$, M = Cr, Mn, Fe, and Co) \cite{moritomo}. 

Investigations on individual compounds or on a sub-family provide vital understandings on the contribution of particular constituents other than M and O, a particular structural symmetry with minor variations in it on the electronic and magnetic structures. However, absence of cross family studies with varied d-orbital occupancy (through variation in M) acts as hindrance to gather the commonalities and thereby to develop phenomenological hypotheses on the relation between the electronic structure and various competing interactions that include bonding and hybridization, crystal field effect,  Hund's rule coupling, on-site Coulomb repulsion. These hypotheses, if envisaged, can assist in predicting the basic electronic and magnetic structure of a given compound in general which is the main objective of this work. 

As a first attempt, we have carried out extensive DFT calculations on three families of antiferromagnetic and insulating TMOs, {\it viz.}, monoxides (MO), spinels (ZnM$_2$O$_4$) and lithium based olivine phosphates (LiMPO$_4$); M = V, Cr, Mn, Fe, Co and Ni. Each members of these families have MO$_6$ complexes. However, for experimentally synthesized MO it is an ideal octahedra (rock-salt structure) \cite{mattheiss, morin, roth}, for ZnM$_2$O$_4$ it is a pseudo-octahedra \cite{mitchell} and for LiMPO$_4$ it is completely asymmetric \cite{ajit}. Similarly while in monoxide, the electronic structure is governed by the M-O complex, spinels have an additional member in the form of Zn and olivine phosphates have two additional members Li and P with the latter making a polyanionic complexes (PO$_4$)$^{3-}$. Though experimentally synthesized MOs and Zn spinels are relatively well studied, the olivine phosphates are not examined enough in the context of structural asymmetry. This can be observed from the fact that in most of the studies \cite{tang, shi1, zaghib} it is still assumed that these compounds have MO$_6$ octahedra and therefore the d-states split into t$_{2g}$ and e$_g$ states. On the contrary, as we will see later, the d-states are completely non-degenerate which is the primary cause of these compounds being insulating. 

These carefully chosen three families allow us to construct an empirical hypothesis to relate the d-band filling (d$^n$), crystal field split ($\Delta_{cr}$), spin-exchange split ($\Delta_{ex}$), and on-site Coulomb repulsion (U) with the electronic and magnetic properties of any TMO with MO$_6$ complexes. To develop the hypothesis, DFT calculations on many hypothetical structures, beyond the ground state, are also carried out. The understanding is further extrapolated, with the support of available literature data, to built similar hypotheses for TMOs with tetrahedral, square-planar, square-antiprismatic as well as asymmetric M-O complexes.

The paper is organized as follows. In  section Section~\ref{sec:structure}, we have presented the structural details of the three prototype families of compounds. Section~\ref{sec:DFT} provides the   the theoretical framework and details of DFT calculations. In Section~\ref{sec:DFT-ES}, we have presented the electronic structures of the three families of compounds and discussed about the effect of various coupling parameters on the electronic structures. The idea generated from the comparison among these coupling parameters led us to develop an empirical hypotheses about the electronic structure of various 3d TMOs which are discussed in Section~\ref{sec:schematics}. Finally, Section~\ref{sec:conclusion} presents summary and future prospects. 

\section{Structure of the Prototype Compounds}
\label{sec:structure}

\begin{figure*}
\centering
\includegraphics[scale = 0.45]{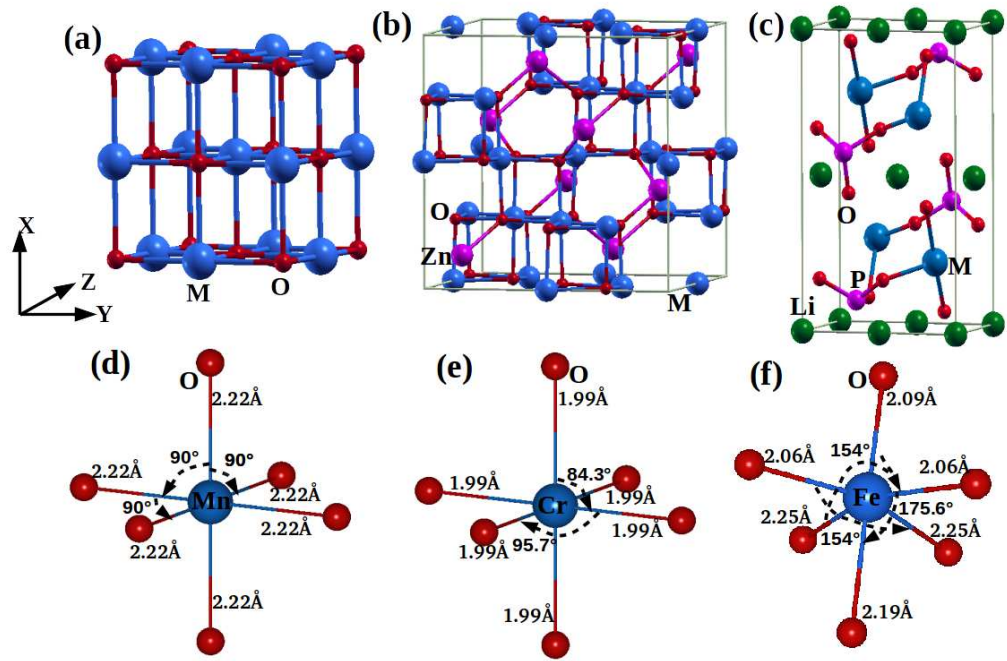}
\caption{Crystal structure of (a) Transition metal monoxide (MO), (b) Spinel (ZnM$_2$O$_4$), and (c) Olivine Phosphate (LiMPO$_4$). Monoxide and Spinel compounds crystallize in fcc phase, whereas Olivine phosphate are in orthorhombic phase. (d-f) MO$_6$ complex for monoxide, spinel and  olivine phosphate, respectively. The bond lengths and bond angles of the MO complexes, for one system from each family  i.e., MnO (d), ZnCr$_2$O$_4$ (e) and LiFePO$_4$ (f), are shown to demonstrate the structural dissimilarities. In case of MnO, all Mn-O bond lengths are same (2.22 \AA{}) and O-Mn-O bond angles are equal to 90$^\circ$, therefore, it forms a perfect octahedra. In case of ZnCr$_2$O$_4$ spinel, the Cr-O bond lengths are all same (1.99 \AA{}) but $\angle$ O-Cr-O $\neq 90 ^\circ$, (out of total twelve O-Cr-O bond angles, six in the positive and negative octants have  84.3$^\circ$ and the rest have 95.7$^\circ$). Spinel MO$_6$ complexes are not perfect octahedra, due to the deviation in the bond angle, therefore spinel family have lower symmetry than that of the monoxides. Considering the case of LiFePO$_4$, which have neither equal Fe-O bond lengths nor the $\angle$ O-Fe-O are equal to 90$^\circ$, therefore in this family MO$_6$ complex is highly asymmetric.}
\label{fig:crystal}
\end{figure*}

We have carried out DFT calculations on a selective set of transition metal oxides, {\it viz.}, monoxides (MO), spinel compounds (ZnM$_2$O$_4$) and olivine phosphates (LiMPO$_4$) to demonstrate and develop a hypothesis connecting  $\Delta_{cr}$, $\Delta_{ex}$, $U$ and electron hopping strength ($t$) with the electronic and magnetic properties of TMOs. The crystal structure of these compounds are given in Fig.~\ref{fig:crystal}.

{\em Transition Metal Monoxides} (MO, M = V, Cr, Mn, Fe, Co and Ni) : The crystal structure of this family of compounds is identical to that of the rocksalt, see Fig.~\ref{fig:crystal}(a), with space group $Fm\bar{3}m$ (No.225). The Wyckoff positions of M and O atoms are 4a : (0,0,0) and 4b : ($\frac{1}{2}, \frac{1}{2}, \frac{1}{2}$) respectively. The experimental lattice parameters considered to carry out the calculations, are given in Table~\ref{tab:tmmo}. In this series of compounds, the transition metal (M) atoms are divalent and are octahedrally coordinated with oxygen atoms. As experimental synthesis of CrO in bulk form is yet to be realized, we have considered a hypothetical cubic CrO with an optimized lattice parameter of 4.140 \AA{}. The monoxide structure possesses maximum symmetry operations that a transition metal oxide can have with the highlight of O$_h$ octahedral symmetry of the M-O complex.

\begin{table} 
\centering
\caption{The experimental lattice parameter $a$ of transition metal monoxides (MO) used in this calculation.}
\begin{ruledtabular}
\begin{tabular}{ccccccc}
Transition Metal (M)& V & Cr & Mn & Fe & Co & Ni \\ \hline
$a$ (\AA{}) & 4.090 & 4.140* & 4.435 & 4.334 & 4.254 & 4.173  \\
References & \cite{koiller} &  & \cite{mattheiss} & \cite{mccammon} & \cite{carey} & \cite{carey} \\ \end{tabular}
\label{tab:tmmo}
\end{ruledtabular}
* For CrO, theoretically optimized lattice parameter is given as Cr does not experimentally form bulk monoxide.
\end{table}

{\em Spinel Compounds} (ZnM$_2$O$_4$, M = V, Cr, Mn, Fe, Co and Ni) : This family crystallizes in face centered cubic structure (Fig.~\ref{fig:crystal}(b)) with space group $fd\bar{3}m$ (No. 227), with an exception to ZnMn$_2$O$_4$. ZnMn$_2$O$_4$ exists as the four formula unit tetragonal phase having $c/a = 1.62$ at room temperature and makes a transition to cubic phase 1323 K \cite{irani}. To compare the electronic properties as a function 3d$^n$ in  this series, we have considered all the spinels in their cubic phase. For the cubic ZnMn$_2$O$_4$, we have considered the structural parameters of the nearest member ZnCr$_2$O$_4$ to carry out the DFT calculations. 

In the cubic phase, Wyckoff positions of Zn, M and O atoms are 8a: ($\frac{1}{8}, \frac{1}{8}, \frac{1}{8})$, $16d: (\frac{1}{2}, \frac{1}{2}, \frac{1}{2}$) and 32e : ($x, x, x$) respectively. In an ideal spinel $x=0.25$, but experimentally it is obtained that $x>0.25$, i.e., the ideal structure is distorted along $\langle 111͘\rangle$ directions \cite{soliman}. In the spinel, as $x$ increases above the ideal value, anions move away from the tetrahedrally coordinated Zn-cations along $\langle 111͘\rangle$ direction.
The divalent Zn have tetrahedral and trivalent M have octahedral symmetry with the O atoms. We have carried out these calculations using the experimental values of lattice constants ($a$) and internal parameter $x$ as listed in Table~\ref{tab:spinel}. 

\begin{table} 
\centering
\caption{The lattice parameter ($a$) and the oxygen atomic position ($x$) of spinel compounds (ZnM$_2$O$_4$) used in this calculation.}
\begin{ruledtabular}
\begin{tabular}{ccccccc}
Transition Metal (M) & V & Cr & Mn & Fe & Co &  Ni \\ \hline
$a$ (\AA{})& 8.403 & 8.327 & 8.327 & 8.436 & 8.104 & 8.396\\   
$x$		   & 0.260 & 0.262 & 0.262 & 0.258 & 0.263 & 0.272 \\ 
References & \cite{reehuis} & \cite{sawada} &  & \cite{gomes} & \cite{dekker} &  \\
\end{tabular}
\label{tab:spinel}
\end{ruledtabular}
\end{table}

{\em Olivine Phosphate} (LiMPO$_4$, M = V, Cr, Mn, Fe, Co and Ni) : The olivine phosphate family crystallizes in the orthorhombic lattice system (Fig.~\ref{fig:crystal}(c)) with space group Pnma (No. 62). It has three non-equivalent O sites. O$_1$ and O$_2$ occupy 4c site, O$_3$ occupies 8d site, M and P occupy 4c site. The Wyckoff positions of Li is 4a (0,0,0). While monoxides have regular MO$_6$ octahedra (Fig.~\ref{fig:crystal}(d)) and spinels have pseudo-octahedra (Fig.~\ref{fig:crystal}(e)), the olivine phosphates have highly asymmetric MO$_6$ \cite{ajit} complexes as can be seen from Fig.~\ref{fig:crystal}(f). It has been found that such a symmetry breakdown is triggered by the formation of a stable PO$_4$ tetrahedra \cite{ajit}. Therefore, these three selective families of compounds can be treated as prototype to examine the evolution of electronic and magnetic structure of TMOs with symmetry lowering. Since V and Cr based olivine phosphates are yet to be experimentally synthesized, we have examined the theoretically optimized structures. The structural parameters for the Olivine family are listed in Tables~\ref{tab:olivine} and~\ref{tab:olivine1}.

\begin{table} 
\centering
\caption{Theoretically optimized structural parameters for LiMPO$_4$.}
\begin{ruledtabular}
\begin{tabular}{ccccccc}
Transition & V & Cr & Mn & Fe & Co &  Ni \\ 
 Metal (M) & & & & & & \\ \hline
$a$ (\AA{}) & 10.291 & 10.253 & 10.502 & 10.441 & 10.335 & 10.156  \\ 
$b$ (\AA{}) & 5.996  & 6.287  & 6.142  & 6.083  & 5.998  & 5.935   \\ 
$c$ (\AA{}) & 4.725  & 4.759  & 4.770  & 4.748  & 4.747  & 4.730  \\ 
\end{tabular}
\label{tab:olivine}
\end{ruledtabular}
\end{table}

\begin{table} 
\centering
\caption{The optimized atomic positions of Olivine Phosphate (LiMPO$_4$).}
\begin{ruledtabular}
\begin{tabular}{ccccccccccccccccccccc}
Transition && Li & M & P & O$_1$ & O$_2$ & O$_3$  \\ 
Metal (M)&  & (4a) & (4c) & (4c) & (4c)& (4c) & (8d) \\ \hline
 
  &x & 0.0 & 0.272 & 0.087 & 0.090 & 0.448 & 0.158 \\ 
 V & y & 0.0 & 0.250 & 0.250 & 0.250 & 0.250 & 0.044 \\ 
  &z & 0.0 & 0.976 & 0.408 & 0.732 & 0.218 & 0.268 \\\\ 
   
   && 0.0 & 0.285 & 0.092 & 0.101 & 0.448 & 0.158 \\ 
Cr && 0.0 & 0.250 & 0.250 & 0.250 & 0.250 & 0.051 \\ 
   && 0.0 & 0.943 & 0.414 & 0.740 & 0.195 & 0.277 \\\\
 
   && 0.0 & 0.280 & 0.089 & 0.093 & 0.452 & 0.159 \\ 
Mn && 0.0 & 0.250 & 0.250 & 0.250 & 0.250 & 0.048 \\ 
   && 0.0 & 0.969 & 0.407 & 0.730 & 0.219 & 0.274 \\ \\
   
   && 0.0 & 0.276 & 0.091 & 0.095 & 0.449 & 0.162 \\ 
Fe && 0.0 & 0.250 & 0.250 & 0.250 & 0.250 & 0.042 \\ 
   && 0.0 & 0.976 & 0.412 & 0.738 & 0.213 & 0.278 \\ \\
   
   && 0.0 & 0.278 & 0.098 & 0.098 & 0.454 & 0.167 \\ 
Co& & 0.0 & 0.250 & 0.250 & 0.250 & 0.250 & 0.044 \\ 
   && 0.0 & 0.994 & 0.421 & 0.745 & 0.206 & 0.283 \\\\
    
   && 0.0 & 0.275 & 0.094 & 0.098 & 0.451 & 0.166 \\ 
Ni && 0.0 & 0.250 & 0.250 & 0.250 & 0.250 & 0.042 \\ 
   && 0.0 & 0.986 & 0.419 & 0.745 & 0.202 & 0.277 \\ 
\end{tabular}
\label{tab:olivine1}
\end{ruledtabular}
\end{table}

\section{Theoretical Framework and Computational Details}
\label{sec:DFT}

In the last few decades several methods and techniques have been developed to accurately solve the many-electron Schrodinger's equation. Particularly to address the correlation effect, very relevant for 3d transition metal oxides, DFT+U \cite{anisimov2}, dynamical mean-field theory (DMFT) \cite{kotliar} and Green's function based GW method \cite{hedin} are developed in recent past to accurately calculate the ground state electronic structure of solids. However, mean-field based DFT+U formalism has remained the most comprehensive and computationally efficient tool for developing phenomenology involving the interplay between spin, charge, orbital and correlations. Such phenomenology are vital in define the electron behavior in crystalline solids. The DFT+U formalism is based on two independent frame works. First one involves the mapping of many-electrons system to a non-interacting quasi particle system with identical particle density  which results in the evaluation of Kohn-Sham states as follows:

\begin{eqnarray}
\begin{split}
E_{KS}=& T_s[n]+\int d^3 r V_{ext}({\bf r}) n({\bf r}) + E_{Hartree}[n]\\
+&E_{II} + E_{XC}[n]
\end{split}
\end{eqnarray}

Here, $T_s$ is the quasi particle kinetic energy. The second term couples the particle motion to the periodic potential offered by the ions as well as any other potential either applied or due to defects and impurities. The third term sums the electrostatic interactions between the valence electrons.

The term $E_{II}$ (in Eq. (1)) represents the ion-ion interaction and therefore acts as a reference potential with no relevance to the electron behavior in the solid. The last term is the exchange-correlation energy that adds up any other interaction not represented by the quasi particle: 

\begin{equation}
E_{XC}[n] = \langle \hat{T} \rangle - T_s[n] + \langle \hat{V}_{int} \rangle - E_{Hartree}[n]
\end{equation}

As the exact expression of the exchange-correlation functional is not known, several approximations such as LDA, GGA are employed and there lies the major disadvantage of DFT in estimating the exact band gap, lattice volume etc. 

The second framework of DFT+U formalism is the inclusion of Hubbard $U$ which adequately accounts the on-site Coulomb repulsion that was missing E$_{XC}$. The parametric $U$ is instrumental in developing several phenomena like band insulator, Mott insulator, charge transfer insulator, mixed insulator and correlated metals. 

To carry out the DFT calculations in the present paper, we have employed the pseudopotential approximation, which basically approximate the diverging V$_{ext}$ in the immediate neighborhood of the nuclei, and used the PBE-GGA exchange correlation functional \cite{pbe}. The Kohn-Sham eigenstates are self consistently evaluated using plane wave basis sets as implemented in Quantum ESPRESSO \cite{giannozzi}. In all of the calculations ultra-soft pseudopotentials are considered. The kinetic energy cut-off for the plane waves and charge densities are taken as 30 Ry and 250 Ry, respectively. We have also incorporated the Hubbard U to account the strong correlation effect. The Brillouin zone integration for the self-consistent calculation is carried out with a Monkhorst-Pack grid. We find that a k-mesh of $16\times16\times16$ (417 irreducible points) for monoxides, $6\times6\times6$ (46 irreducible points) for spinels, and $6\times10\times12$ (456 irreducible points) for olivine phosphates is sufficient to converge the total energy and charge densities. 

The ground state results obtained from the first principles electronic structure calculations can be directly mapped to a general Hamiltonian as follows to gain further insight in order to develop hypotheses connecting the electronic structure with several competing parameters. 

\begin{eqnarray}
\begin{split}
H = & \sum_{\substack{i j \alpha \beta \sigma\\i\neq j}}t_{i j \alpha \beta} c_{i \alpha \sigma}^\dagger c_{j \beta \sigma} +h.c.\\
+& \sum_{i \alpha \beta \sigma} \frac{(\Delta_{cr})_{i \alpha \beta}}{2} \{c_{i \alpha \sigma}^\dagger c_{i \alpha \sigma} - c_{i \beta \sigma}^\dagger c_{i \beta \sigma}\} \\
+& \sum_{i\alpha}(\Delta_{ex})_\alpha\{c_{i \alpha \uparrow}^\dagger c_{i \alpha \uparrow} - c_{i \alpha, \downarrow}^\dagger c_{i \alpha, \downarrow}\}\\
+& U_1 \sum_{i \alpha} n_{i \alpha \uparrow}n_{i \alpha \downarrow}+ U_2 \sum_{i  \alpha \beta \sigma} n_{i \alpha \sigma} n_{i \beta \sigma} \\
+& U_3 \sum_{i \alpha \beta} n_{i \alpha \uparrow} n_{i \beta \downarrow} + \sum_{ij} J_{ij}^{mag} \bm{S_i} \cdot \bm{S_j}
\end{split}
\end{eqnarray}

Here, $c^\dagger$ and $c$ are creation and annihilation operators and $i$, $\alpha$ and $\sigma$ denote site, orbital and spin indices respectively. 
The first term is the electron hopping energy with hopping integral  $t$ and directly related to the kinetic energy of the system as projected in Eq. (3) and (5). The bandwidth in the independent electron approximation (i.e., in the absence of external strong correlation measure such as Hubbard U) provides a quantitative measure of $t$.
The second term summarily presented the crystal field effect which in DFT is incorporated through V$_{ext}$ (see Eq. (1)). 

\begin{equation}
(\Delta_{cr})_{\alpha \beta} = \Delta_{cr} + \Delta_{cr}'
\end{equation}

\begin{figure}
\centering
\includegraphics[scale=0.5]{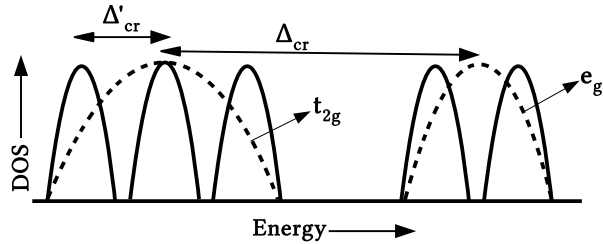}
\caption{In an octahedral complex, due to crystal field effect ($\Delta_{cr}$), the five fold d-states splits into three-fold $t_{2g}$ and two-fold $e_g$ degenerate states. The lowering in the octahedral symmetry introduces an asymmetric crystal field which  ($\Delta_{cr}'$) which further splits the $t_{2g}$ and $e_g$ states into five non-degenerate states.}
\label{fig:cf}
\end{figure}

In transition metal oxides O ligands exert such a field on the electrons occupying the M-d states which in turn splits the five-fold degenerate states to few multi-fold degenerate states depending on the symmetry of the crystal field. For example in case of an MO$_6$ octahedral complex the d-states are split into three-fold t$_{2g}$ and two fold e$_g$ states as shown in Fig.~\ref{fig:cf}. Further lowering in the crystal symmetry introduce an additional splitting $\Delta_{cr}'$  as demonstrated in Fig.~\ref{fig:cf}. 

The strength of the spin-polarization (or magnetization) is measured by the spin-exchange split ($\Delta_{ex}$), as expressed in the third term of the Hamiltonian (Eq. (6)). In atoms this is governed by the Hund's rule coupling. In solids, beside Hund's rule coupling, $\Delta_{ex}$ is governed by charge state of the transition metal ion and the crystal field. The fourth, fifth and sixth terms together represent the on-site Coulomb repulsion. The repulsion strength U$_1$ is between states with same orbital and opposite spins, U$_2$ is between states with same spins and different orbitals and U$_3$ is between states with different orbitals and opposite spins. 
The last term in Eq. (6) is a Heisenberg Spin Hamiltonian representing the coupling between the neighboring spins to determine the ground state magnetic ordering of the system.  Generally with the aid of DFT calculations, the total energies of several possible spin-arrangements are compared to determine the ground state spin-arrangement. In this work, we have not focused on the magnetic ordering which in principle is very much compound specific. However, the spin-polarized electronic structure of the M-O complex of the prototype compounds are examined in great details.

\section{Electronic Structure of the Prototype TMOs}
\label{sec:DFT-ES}

In this section we shall examine the electronic structure of the prototype antiferromagnetic and insulating TMOs, {\it viz.}, monoxides, spinels and Olivine phosphates, as obtained from the DFT calculations. The structural details of these compounds are already presented in Section~\ref{sec:structure}. From the DFT results, we will estimate the {\it ab initio} values of the coupling parameters $t$, $\Delta_{cr}$, $\Delta^{\prime}_{cr}$ and $\Delta_{ex}$ in order to devise the desired hypothesis in the later section. 

\subsection{Transition Metal Monoxides: MO}

\begin{figure}
\centering
\hspace*{-1cm}\includegraphics[scale=0.9]{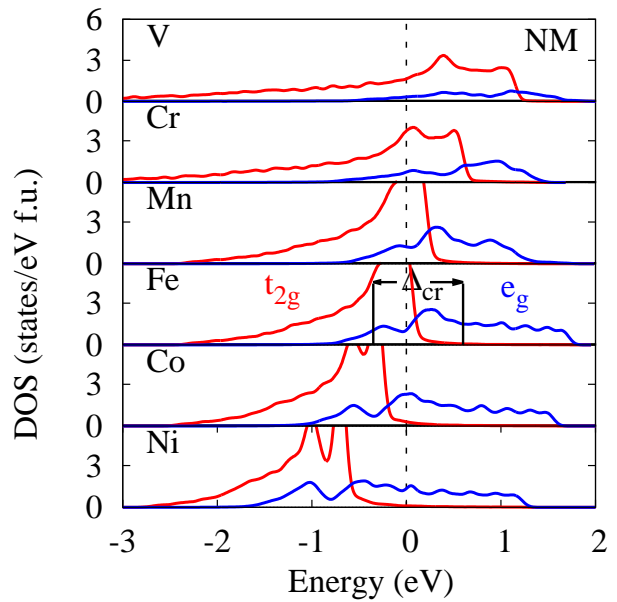}
\caption{ Non-magnetic $t_{2g}$ and $e_{g}$ partial DOS for MO. (M = V, Cr, Mn, Fe, Co and Ni). $E_F$ is set to zero.}
\label{fig:tmmo-crystal} 
\end{figure}

This family of binary compounds with rocksalt structure and ideal MO$_6$ octahedra, as shown in Fig.~\ref{fig:crystal}(a) and (d), satisfies maximum point group symmetry operations (O$_h$)  among the TMOs. In addition, unlike the other TMOs, in the absence of any third element, electronic structure is purely governed by the interactions between the O-p and M-d states. The octahedral crystal field splits five fold degenerate d-states into triply degenerate $t_{2g}$ (d$_{xy}$, d$_{yz}$ and d$_{xz}$) states lying lower in energy and doubly degenerate $e_g$ (d$_{x^2-y^2}$ and d$_{3z^2-r^2}$) states lying higher in energy. Since the transition metals are divalent in monoxides, the d-occupancy ($d^n$) in the present study varies from $n=3$ (for VO) to $n=8$ for (NiO).

\begin{figure}
\centering
\includegraphics[scale=0.5]{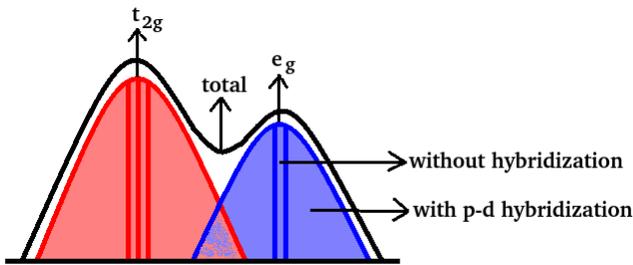}
\caption{Schematic diagram representing the overlapping of the $t_{2g}$ and $e_{g}$ bands due to strong  p-d and d-d covalent hybridizations.}
\label{fig:d-state} 
\end{figure}

To examine the competition between the crystal field effect, which is a charge phenomena, and the electron hopping strength, measured from the band width, in  Fig.~\ref{fig:tmmo-crystal} we have plotted the $t_{2g}$ (red) and $e_g$(blue)  DOS in the non-magnetic (NM) state. In the absence of $p-d$ and $d-d$ hybridization, one would have seen a clear splitting between $t_{2g}$ and $e_g$ states as schematically illustrated in Fig.~\ref{fig:d-state}. However, with these hybridizations, $t_{2g}$ and $e_g$ states spread and might overlap in the energy spectrum to alter the $t_{2g}$ and $e_g$ occupancies. In a localized picture, the e$_g$ states would have been remained empty for d$^{3-6}$ systems, in the non-magnetic configuration (see Fig.~\ref{fig:tmmo-crystal}). But the occupancy, calculated by integrating the orbital resolved DOS up to E$_F$, as listed in Table~\ref{tab:tmmo-d-occupancy} shows that, for these states, the e$_g$ occupancy is substantial ($\geq 0.36 $) in the non-magnetic case. The crystal field splitting, measured by estimating the separation between the band center of t$_{2g}$ and e$_g$ states, remained almost same ($\sim$ 0.8 eV) across the members of the monoxide family (see Table~\ref{tab:tmmo-energy}).

\begin{table*}
\centering
\caption{The t$_{2g}$ and e$_g$ occupancies of MO for different magnetic configurations. Here, $\uparrow$ and $\downarrow$ represent the spin-up and spin-down states respectively.}
\begin{ruledtabular} 
\begin{tabular}{ccccccccccccccccccc}
Transition & &  \multicolumn{2}{c}{NM}  & \multicolumn{4}{c}{FM} & \multicolumn{4}{c}{AFM} & \multicolumn{4}{c}{AFM+U} \\ 
Metal (M)&\multicolumn{1}{c}{\raisebox{1.5ex}[0pt]{ $d^n$ }} & $t_{2g}$ & $e_g$ & $t_{2g} \uparrow$ &$e_g\uparrow$ & $t_{2g} \downarrow$ & $e_g\downarrow$& $t_{2g} \uparrow$& $e_g\uparrow$&$t_{2g}\downarrow$ & $e_g\downarrow$& $t_{2g} \uparrow$ &  $e_g\uparrow$&$t_{2g}\downarrow$ & $e_g\downarrow$\\ \hline
V & $d^3$ & 2.554 & 0.446& 1.577 &0.376 & 0.996& 0.051& 2.650 & 0.037& 0.235 & 0.078 & 3 & 0 & 0 & 0\\
Cr & $d^4$ & 3.641 & 0.359 & 3 & 0.649 & 0.345 & 0.006 & 3 & 0.973& 0 & 0 & 3 & 1.078 & 0 & 0 \\
Mn & $d^5$ & 4.068 & 0.932 & 3 & 2 & 0.490 & 0 & 3 & 2 & 0 & 0 & 3 & 2 & 0 & 0  \\
Fe & $d^6$ & 5.251 & 0.749 & 3 & 2 & 0.980 & 0 & 3 & 2 & 0.925 & 0 & 3 & 2 & 1.069 & 0\\
Co & $d^7$ & 5.643 & 1.357 & 3 & 2 & 1.917 & 0.083 & 3 & 2 & 2.043 & 0 & 3 & 2 & 1.858 & 0\\
Ni & $d^8$ & 6 & 2.415 & 3 & 1.906 & 3 & 0.094& 3 & 2 & 3 & 0 & 3 & 2 & 3 & 0\\
\end{tabular}
\label{tab:tmmo-d-occupancy}
\end{ruledtabular}
\end{table*}

\begin{figure}
\centering
\hspace*{-1cm}\includegraphics[scale=0.9]{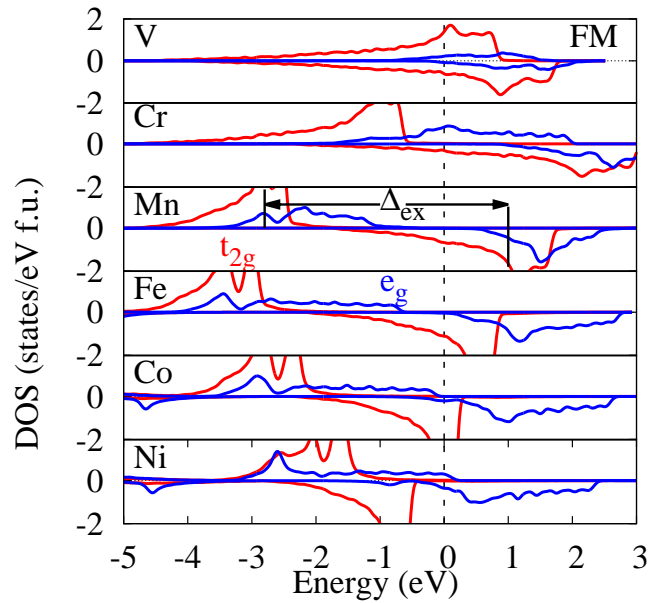}
\caption{Spin-resolved t$_{2g}$ and e$_g$ DOS of MO in the FM configuration. The positive and negative values stand for spin-up and spin-down DOS respectively.}
\label{fig:tmmo-exchange}
\end{figure}

Spin-polarized calculations provides the estimation of spin-exchange split ($\Delta_{ex}$) which according to the Hund's coupling is the energy separation between the band center of the same band in spin-up and spin-down channels. We have extracted the value of  $\Delta_{ex}$ from the ferromagnetic (FM) partial DOS shown in Fig.~\ref{fig:tmmo-exchange} and listed them in Table VI. Along the series with increase in d occupancy, the exchange splitting increases and attends the maximum for MnO. With further increase in occupancy, spin pairing occurs. As a consequence, the $\Delta_{ex}$ reduces and hence the spin-polarization weakens. The FM ordering in the independent electron approximation still allows the monoxides to be metallic, except in the case of MnO due to same reason as illustrated in Fig.~\ref{fig:d-state}.

\begin{figure*} 
\centering
\includegraphics[scale=0.29]{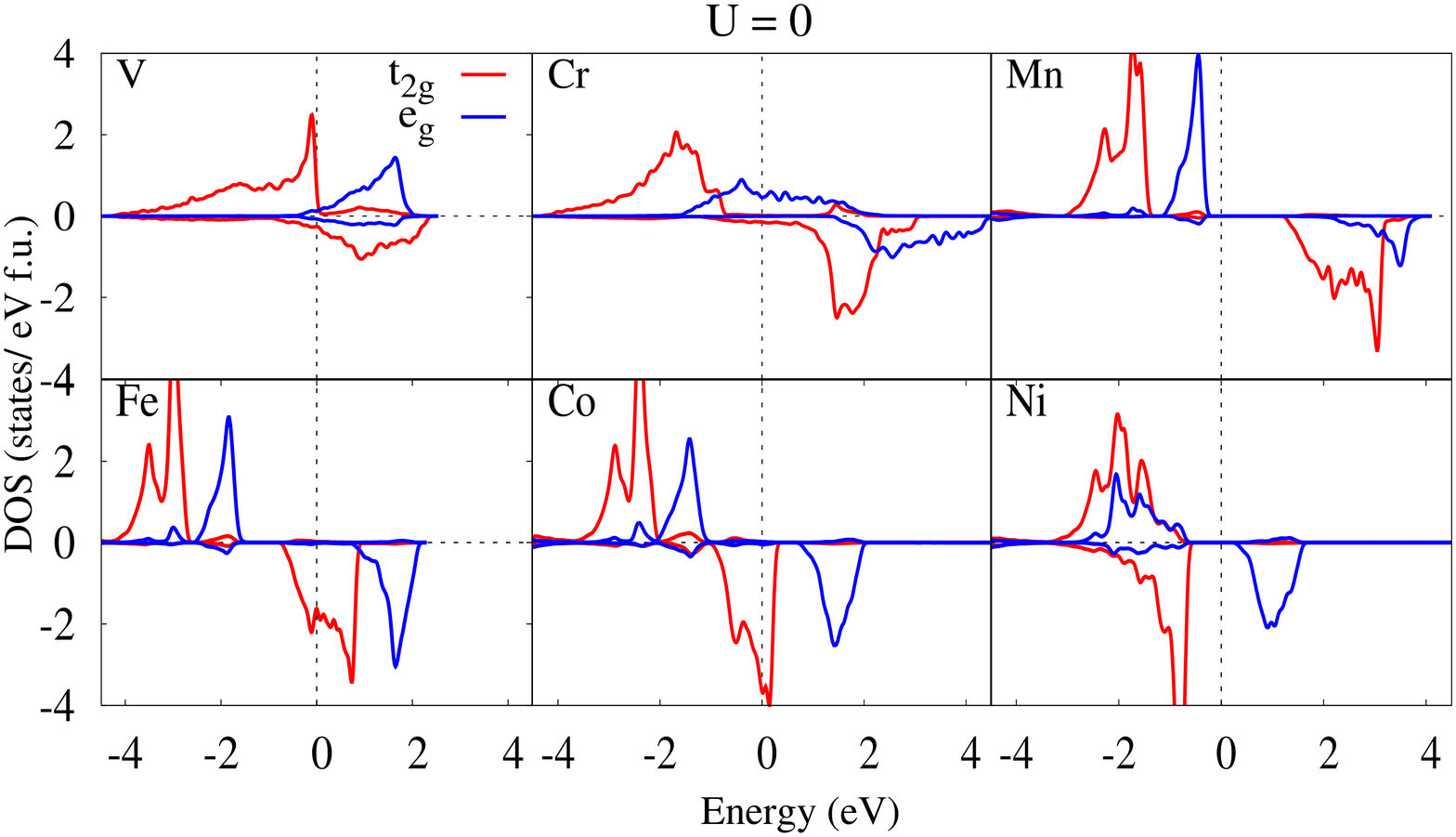}\includegraphics[scale=0.29]{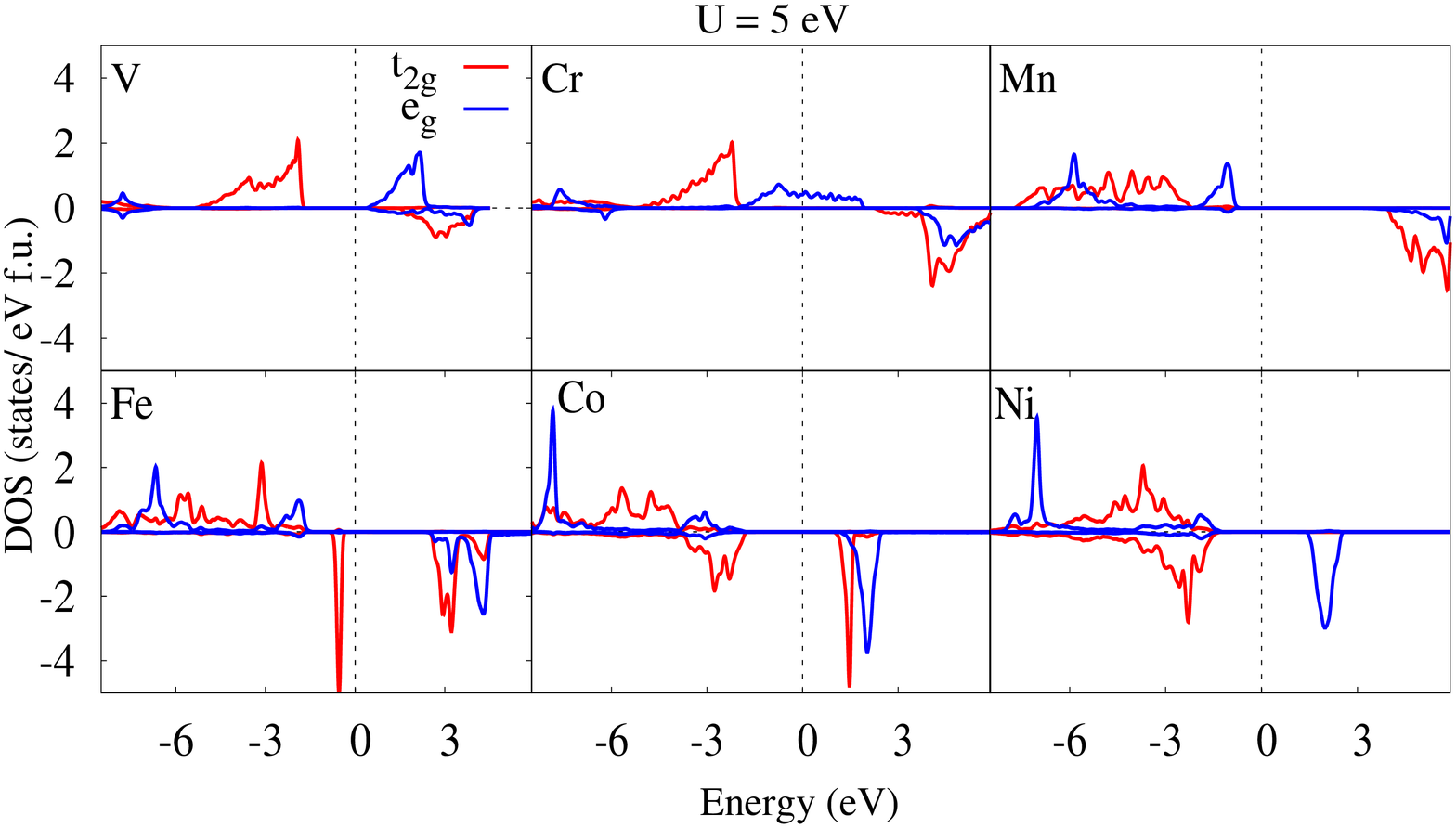}
\caption{GGA and GGA+U calculated spin resolved t$_{2g}$ and e$_g$ DOS for MO in AFM configuration. AFM-II ordering has been considered for all the compounds except CrO. For the latter we find that A-type AFM ordering is far more stable than the AFM-II ordering. Therefore, the A-AFM DOS are displayed here.}
\label{fig:tmmo-afm}
\end{figure*}

In the stable antiferromagnetic configuration, which is AFM-II (A-type antiferromagnetic spin ordering along [111] direction), the covalent hybridizations between two opposite spin sublattices are forbidden. This makes the states localized, as can be seen from Fig.~\ref{fig:tmmo-afm} (left) and therefore the system tends to become insulating. With U = 0, while VO (d$^3$) exhibits a pseudo gap at E$_F$, MnO (d$^5$) and NiO (d$^8$) become insulating. CrO (d$^4$) has half-occupied e$_g$ states in the spin-up channel whereas FeO (d$^6$) and CoO (d$^7$) have partially occupied t$_{2g}$ states in the spin-down channel. The high DOS at E$_F$ make these compounds unstable. The depletion of the DOS at E$_F$ can occurs in two ways. Either through the on-site Coulomb repulsion ($U_2 \sum_{i  \alpha \beta \sigma} n_{i \alpha \sigma} n_{i \beta \sigma}$ of Eq. (6)) which will create two Hubbard subbands, lower Hubbard band (LHB) lying below E$_F$ and upper Hubbard band (UHB) lying above it or through distorting the MO$_6$ octahedra. The distortion lifts the degeneracy of the three-fold t$_{2g}$ or the two-fold e$_g$ states.

\begin{figure}
\centering
\includegraphics[scale=0.4]{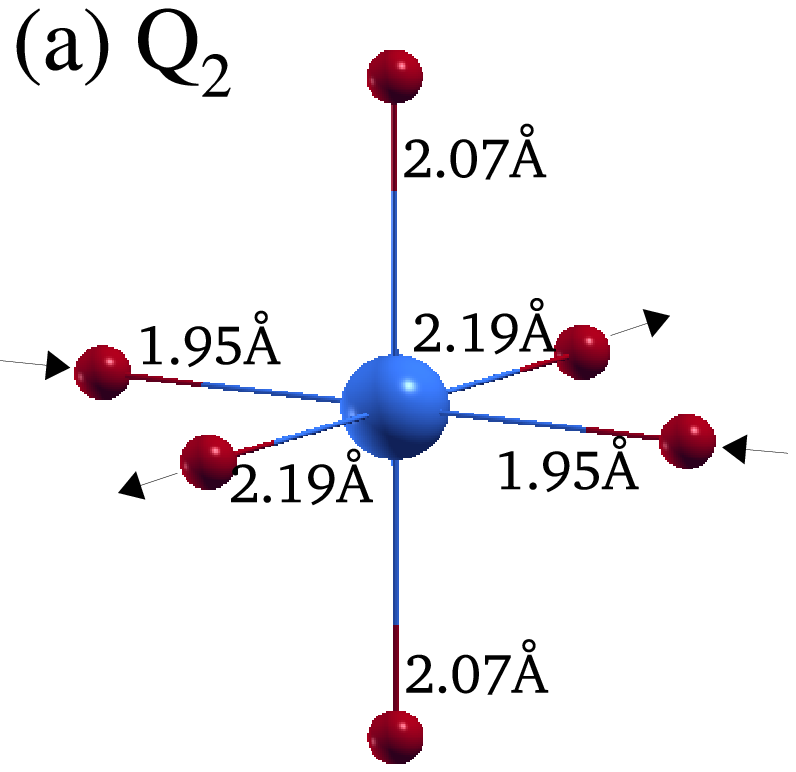}\hspace{0.5cm}\includegraphics[scale=0.4]{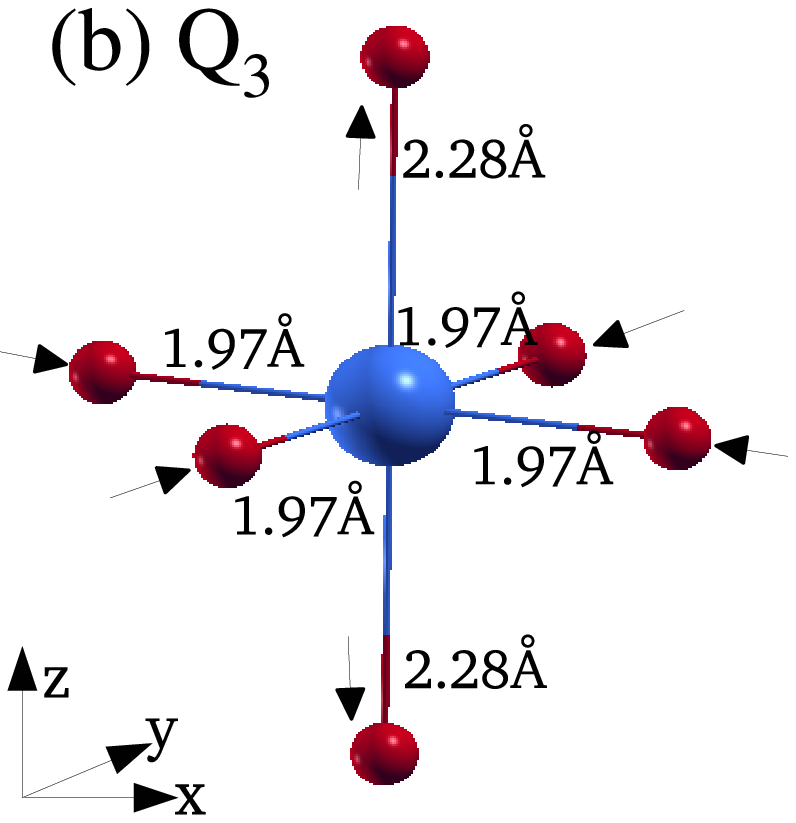}

\hspace*{-0.3cm}\includegraphics[scale =0.46]{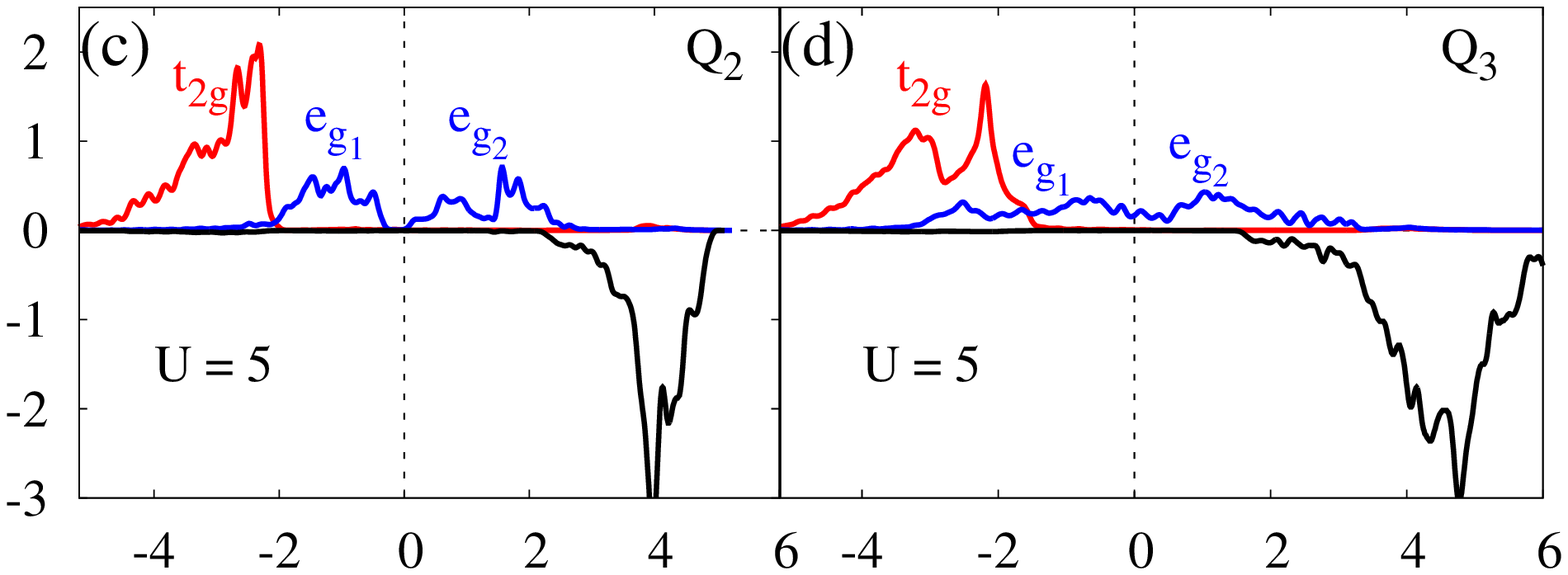}
\caption{JT distortion in the CrO$_6$ octahedra  and its effect on the electronic structure. (a) CrO$_6$ octahedra with Q$_2$ mode of distortion and (b) CrO$_6$ octahedra with Q$_3$ mode of distortion. The corresponding AFM t$_{2g}$ and e$_g$ DOS are shown in (c) and (d) with U = 5 eV.}
\label{fig:cro-jt}
\end{figure}

To examine the strong correlation effect, in Fig.~\ref{fig:tmmo-afm} (right), we have plotted the antiferromagnetic DOS with U = 5 eV. While VO, FeO and CoO have a definite gap at E$_F$, the band gap of NiO and MnO are amplified. However, CrO still has a metallic state. The virtually synthesized compound CrO has a d$^4$ configuration as in the case of LaMnO$_3$. The latter compound undergoes Jahn-Teller (JT) distortion, a combination of Q$_2$ and Q$_3$ modes (see Fig.\ref{fig:cro-jt} (a) and (b)), in order to remove the degeneracy of the e$_g$ states \cite{sashi}. Thereby, it forms a gap with one of the non-degenerate e$_g$ state being occupied and other being empty. To examine if such a distortion will open a gap at E$_F$, in Fig.~\ref{fig:cro-jt}, we have presented the electronic structure with Q$_2$ and Q$_3$ distortions for CrO. We find that within the independent electron approximation (GGA) partially filled e$_g$ states dominate the E$_F$. Only Q$_2$ mode of distortion tend to break the e$_g$ degeneracy. However with inclusion of U (= 5 eV), both Q$_2$ and Q$_3$ distortions tend to open a gap (see Fig.\ref{fig:cro-jt} (c) and (d)). If such a structure can be synthesized experimentally, an A type AFM will be established as in the case of LaMnO$_3$.

\begin{table*}
\centering
\caption{Crystal field  and exchange splitting for transition metal monoxides (MO).}
\begin{ruledtabular}
\begin{tabular}{ccccccccccccc}
& \multicolumn{6}{c}{Crystal Field Splitting (eV)}  & \multicolumn{6}{c}{Exchange Splitting (eV)} \\ 
\multicolumn{1}{c}{\raisebox{0ex}[0pt]{Transition Metal (M)}} & V & Cr & Mn & Fe & Co &  Ni & V & Cr & Mn & Fe & Co &  Ni \\ \hline
This work & 0.66 & 0.92 & 0.62 & 0.84 & 0.85 & 0.79 & 0.71 & 3.20 & 3.92 & 3.72 & 2.64 & 1.15  \\ \\

Other theoretical & & & 0.78$^a$ & 0.80$^a$ &0.76$^a$ & 0.76$^a$ &  &  & 3.81$^b$&  &  & \\ 
 work & & & 1.05$^b$ & 1.14$^b$ &1.16$^b$ & 1.35$^b$ &  &  &  &  &  & \\ \\

Experimental  &  &  & 1.21$^{a,c}$ &  & 1.17$^{a,c}$ &1.13$^{a,e}$ &  &  &  &  &  & \\
 work &  &  &1.25$^{a,d}$ &  &  &1.10$^{a,f}$  &  &  &  &  &  & \\ 
\end{tabular}
\label{tab:tmmo-energy}
\begin{flushleft}
$^a$ Ref.\cite{mattheiss1}, $^b$ Ref.\cite{terakura}, $^c$ Ref.\cite{pratt}, $^d$ Ref. \cite{huffman}, $^e$ Ref. \cite{newman}, $^f$ Ref. \cite{stephens}.
\end{flushleft}
\end{ruledtabular}
\end{table*}

\subsection{Spinel Compounds: ZnM$_2$O$_4$}

\begin{figure}
\centering
\includegraphics[scale=0.8]{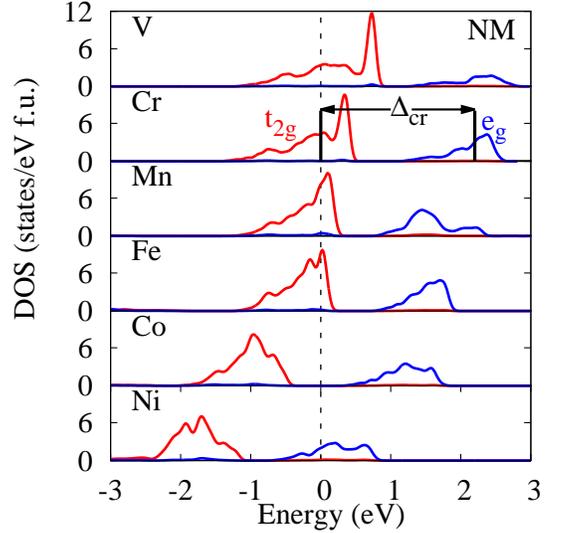}
\caption{The t$_{2g}$ and e$_g$ partial DOS for ZnM$_2$O$_4$ in the NM phase.}
\label{fig:spinel-crystal}
\end{figure}

From the structural prospective, M in normal spinel ZnM$_2$O$_4$ has three major dissimilarities with that of the monoxide. Firstly, in the spinel compound M stabilizes with 3+ charge state, contrary to the 2+ charge state in monoxides. Secondly unlike in monoxides, in spinels, the cubic symmetry of the MO$_6$ complex is broken to yield pseudo-octahedras as shown in Fig.~\ref{fig:crystal}(e). Here the O ions are tilted towards each other in the positive and negative octants of a given octahedra, to make the six  O-M-O bond angles lying in these octants less than $\pi/2$ ($\simeq$ 84$^\circ$). Consequently, the other six O-M-O bond angles become greater than $\pi/2$ ($\simeq$ 96$^\circ$). Furthermore, the M-O bond length in spinel is smaller than that of monoxides by approximately 0.2 \AA{}. As a consequence, the anisotropic covalent interaction with the O-p states as well as the anisotropic crystal field will deform the shape of the d-orbitals which will no longer retain the spherical symmetry.  Hence, it is expected that in the spinel the d electrons will be comparatively more localized than in the case of monoxides. Thirdly, in monoxide each octahedron is connected to six neighboring octahedra  through common O ions which implies the formation of linear M-O-M chains along the crystal axes. However, in ZnM$_2$O$_4$ the linear chains are replaced by zig-zag M-O-M chains with the M-O-M bond angle being close to $\pi/2$. This is significant in the context of indirect magnetic exchange interactions and hence in the magnetic ordering. In fact, our total energy calculations suggest that the spinel compounds have weak AFM coupling compared to the monoxides. 

\begin{figure}
\centering
\includegraphics[scale=0.8]{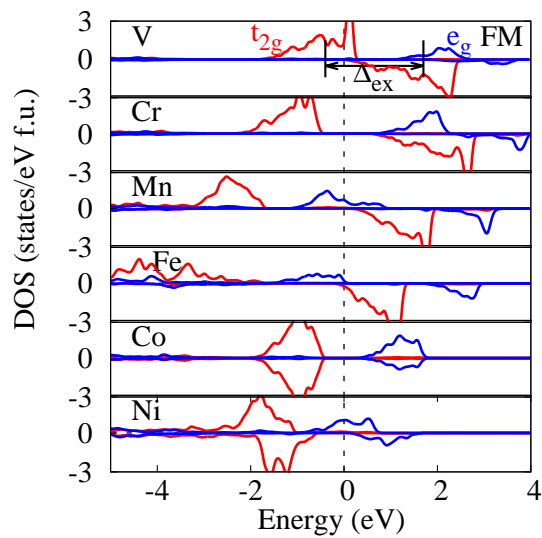}
\caption{Spin resolved t$_{2g}$ and e$_g$ partial DOS of ZnM$_2$O$_4$ in the FM configuration.}
\label{fig:spinel-fm}
\end{figure}

To understand the effect of crystal field split and d-electron hopping, we have plotted the NM (GGA, U = 0) DOS of ZnM$_2$O$_4$ in Fig.~\ref{fig:spinel-crystal}. Compared to the monoxides, the t$_{2g}$ and e$_g$ band width are almost reduced by half, while the crystal field split is almost doubled. The estimated values are listed in Table~\ref{tab:spinel-energy}. The M-d-O-p interaction is reduced by the breakdown of the octahedral symmetry as discussed earlier leading to the localization of the d states. The localized d-electron cloud experiences a greater repulsive field exerted by the O$^{2-}$ ligand which is further amplified by a shorter M-O bond length leading to larger crystal field split. Since Zn$^{2+}$ has d$^{10}$ configuration, it remains far below E$_F$ and do not participate in the chemical bonding.

\begin{table}
\centering
\caption{Crystal field and exchange splitting for Spinels (ZnM$_2$O$_4$), M is 3d transition metal.}
\begin{ruledtabular}
\begin{tabular}{ccccccc}
Transition Metal (M) & V & Cr & Mn & Fe & Co &  Ni \\ \hline
Crystal field Splitting & 1.98 & 2.19 & 2.09 & 1.76 & 2.22 & 2.18 \\ 
Exchange Splitting & 2.16 & 2.66 & 4.40 & 5.40 & 0.00 & 0.50 \\
\end{tabular}
\label{tab:spinel-energy}
\end{ruledtabular}
\end{table}

\begin{figure*}
\centering
\includegraphics[scale=0.3]{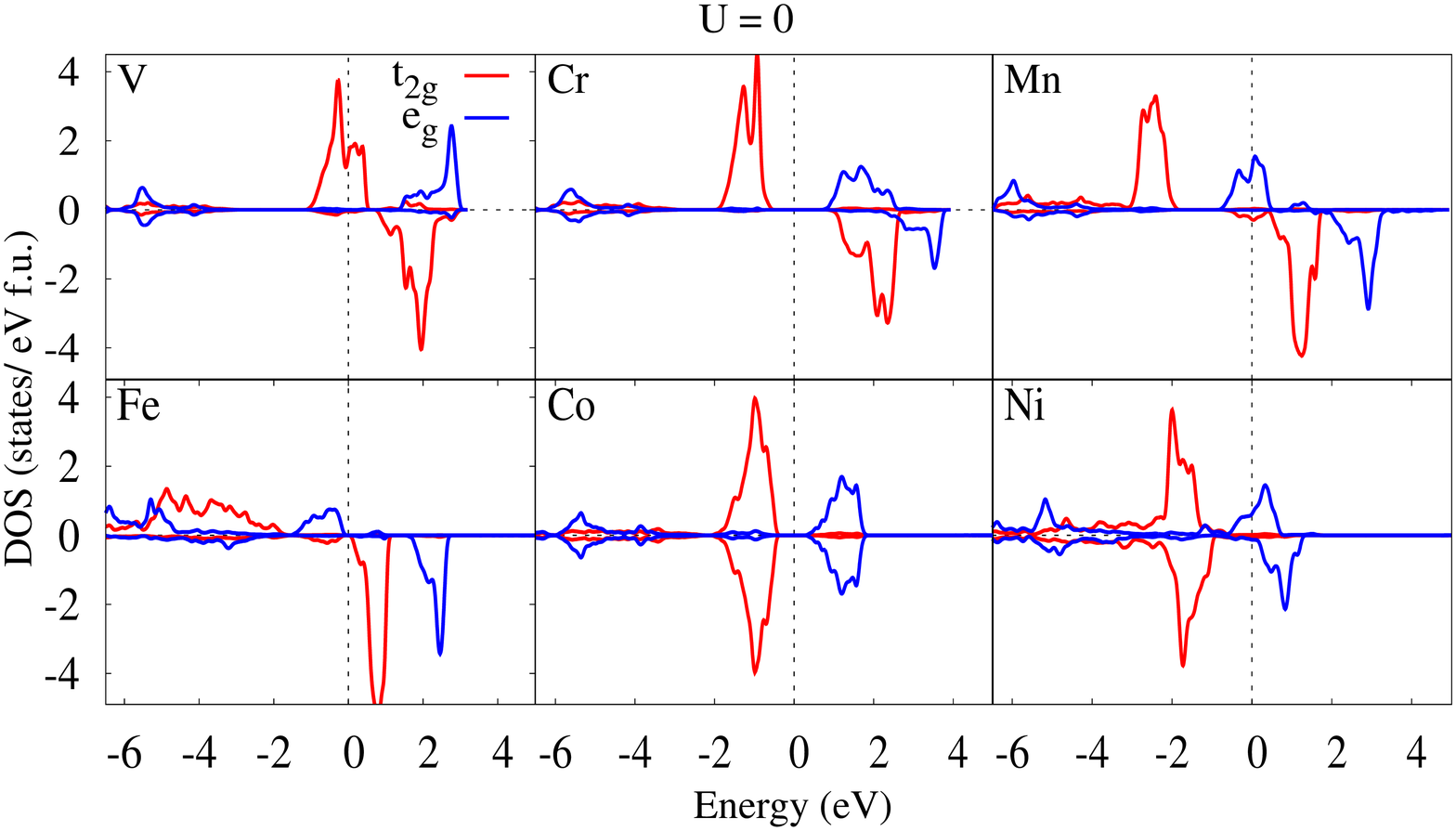}\includegraphics[scale=0.3]{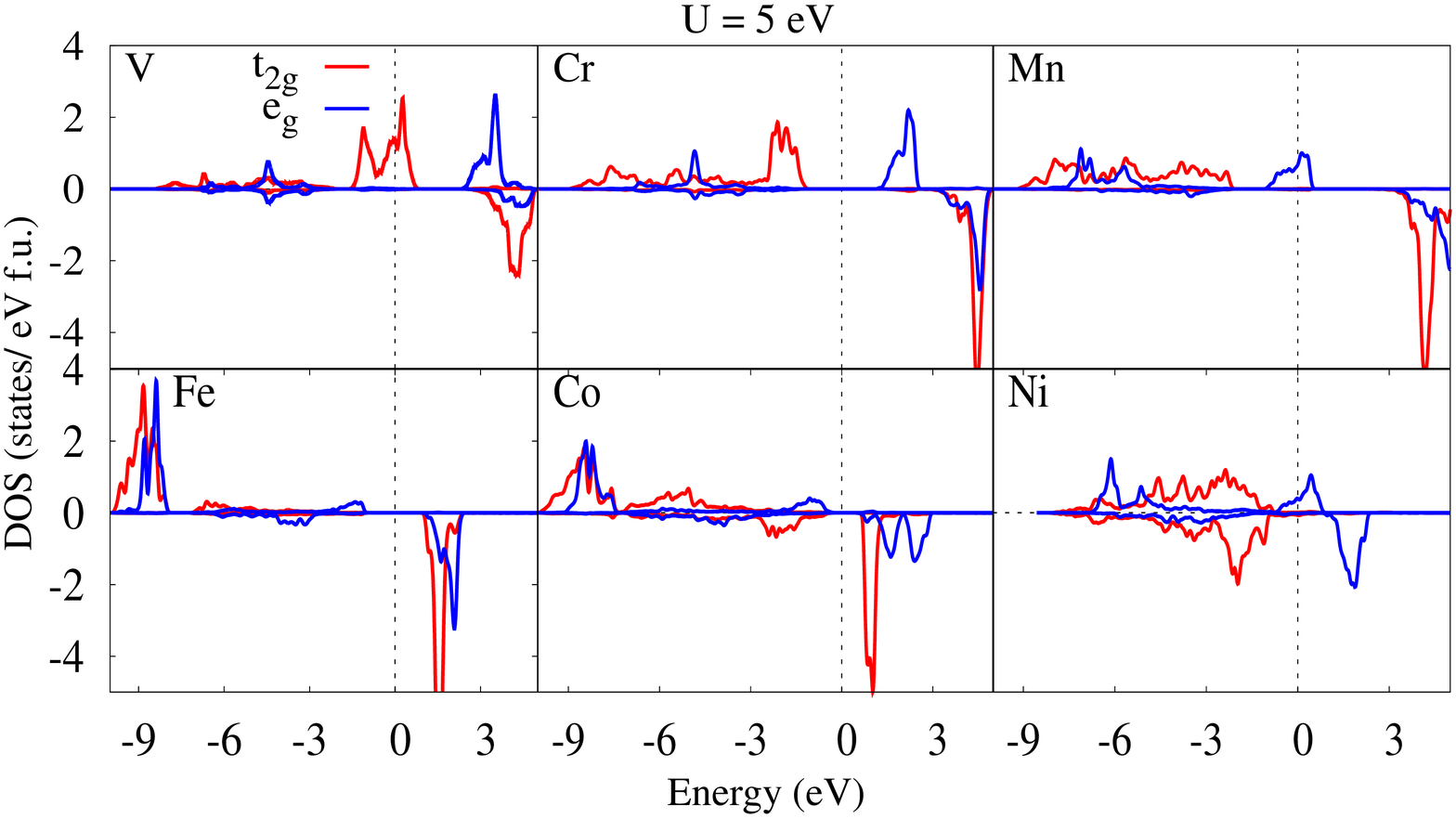}
\caption{GGA and GGA+U calculated spin resolved t$_{2g}$ and e$_g$  DOS for ZnM$_2$O$_4$ in the AFM configuration. For M = V, Mn, and Ni, finite DOS at E$_F$ suggests instability leading to structural distortion (see Fig.~\ref{fig:spinel-distort}).}
\label{fig:spinel-afm}
\end{figure*}

The spin-polarized M-d DOS, calculated with FM ordering, is shown in Fig.~\ref{fig:spinel-fm} to study the spin-split ($\Delta_{ex}$) in the spinel family. As in the case of monoxides,  $\Delta_{ex}$ is largest for the half-filled (ZnFe$_2$O$_4$, d$^5$) configuration (see Table~\ref{tab:spinel-energy}). However, in the independent electron approximation (U = 0), the Co (d$^6$) and Ni (d$^7$) based spinels stabilize in the LS configuration and therefore weak $\Delta_{ex}$ is observed. The values of $\Delta_{ex}$ for each compound are listed in Table~\ref{tab:spinel-energy}. While $\Delta_{ex}$ $>$ $\Delta_{cr}$ for V to Fe, it is reverse for Co and Ni.

As the spin-polarized calculations with FM ordering (Fig.~\ref{fig:spinel-fm}) yield finite DOS at E$_F$ for some of the compounds, namely, d$^2$ (V), d$^4$ (Mn) and d$^7$ (Ni),  these compounds may prefer antiferromagnetic (AFM) insulating phase either through structural distortion or through correlated mechanism (e.g., Mott-Hubbard, charge-transfer etc.) or both. The GGA-AFM DOS plotted in Fig.~\ref{fig:spinel-afm} (left) shows that the width of the t$_{2g}$ and e$_g$ bands are narrowed as there is no electron hopping between the opposite spin-sub lattices. However, the band crossing at E$_F$ for d$^2$, d$^4$ and d$^7$ systems still remains. Fig.~\ref{fig:spinel-afm} (right) suggests that increase in U does not change this basic feature of the electronic structure. With large U, in the case of ZnCo$_2$O$_4$, the LS state changed to HS state. As a consequence, this d$^6$ system become t$^3_{2g}\uparrow$ e$^2_{2g}\uparrow$ t$^1_{2g}\downarrow$ e$^0_{2g}\downarrow$. The strong on-site Coulomb repulsion among the t$_{2g}$ orbitals (U$_2$ and U$_3$ of Eq.(6)) creates a gap at E$_F$ as in the case of FeO (see Fig.~\ref{fig:tmmo-afm}) to make it a correlated insulator.   

\begin{figure}
\centering
\includegraphics[scale=0.3]{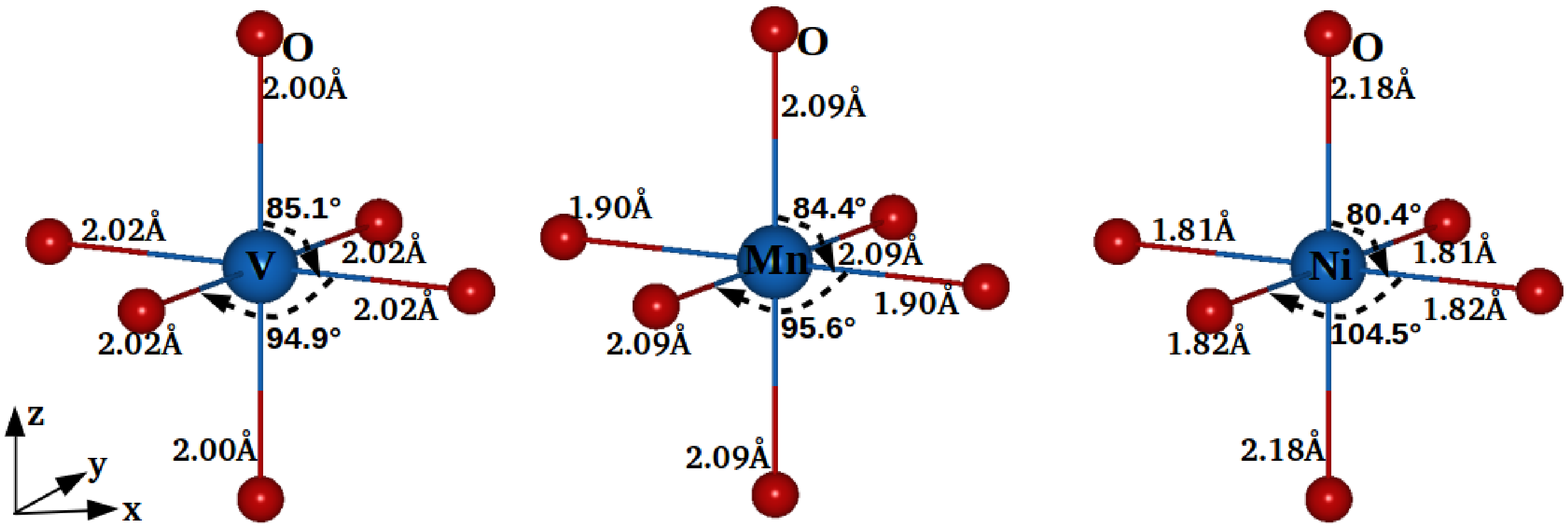}

\hspace*{-0.3cm}\includegraphics[scale=0.29]{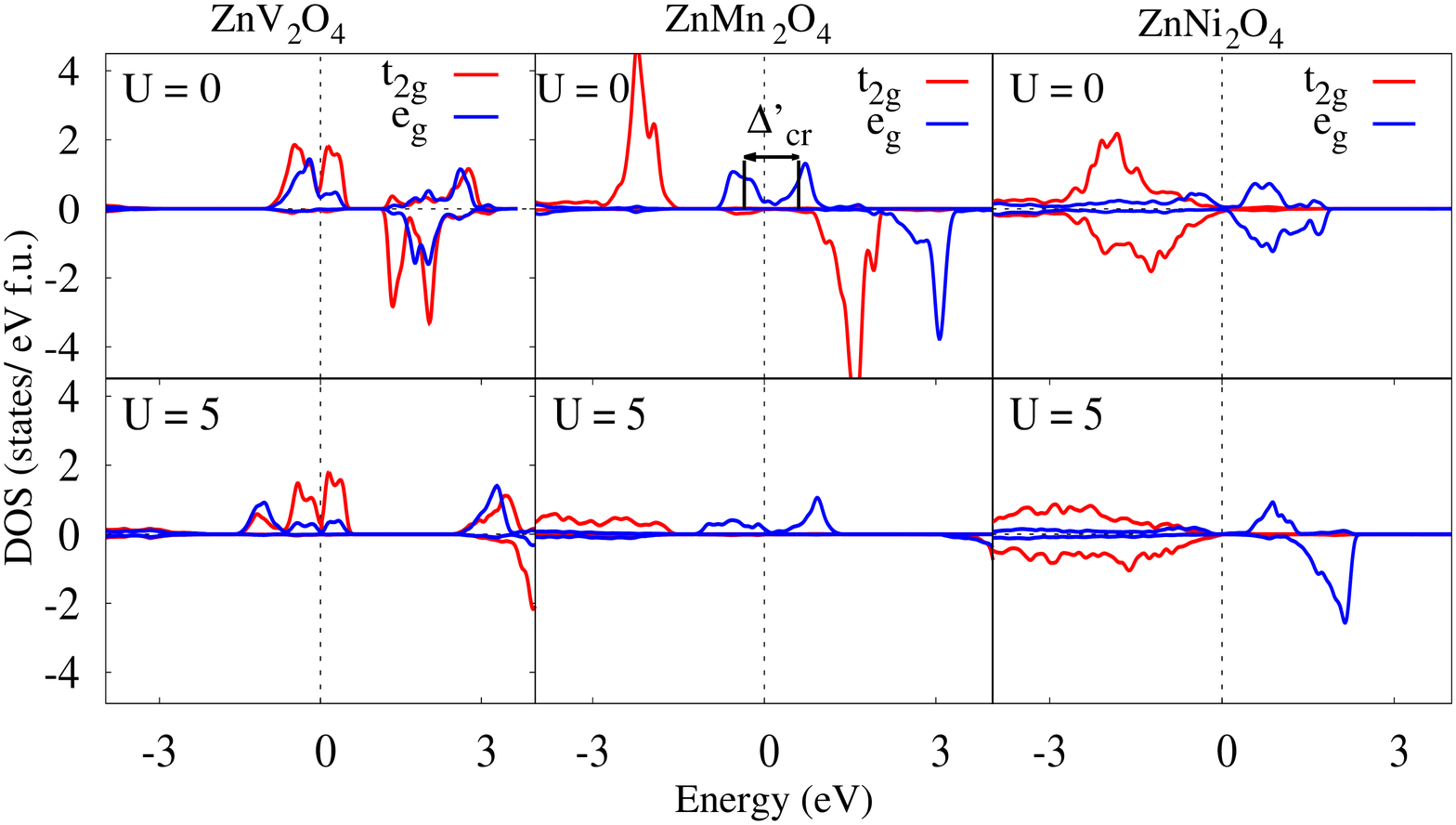}

\caption{JT distortion for ZnM$_2$O$_4$ (M = V, Mn, and Ni) (upper panel) and the corresponding GGA and GGA+U AFM t$_{2g}$ and e$_g$ partial DOS (lower panel) in the vicinity of E$_F$. }
\label{fig:spinel-distort}
\end{figure}

To examine the prospect of splitting of the degenerate states $-$ t$_{2g}\uparrow$ for d$^2$, e$_{g}\uparrow$ for d$^4$ and t$_{2g}\downarrow$ for d$^7$ systems $-$ occupying the Fermi level, the MO$_6$ octahedra are distorted as shown in Fig.~\ref{fig:spinel-distort} (upper panel). The break down of octahedral symmetry introduces the asymmetric crystal field split, $\Delta^{\prime}_{cr}$, and remove the degeneracy to open a gap at E$_F$. While, in ZnV$_2$O$_4$, both strong correlation effect and crystal field distortion are instrumental in opening the gap, for ZnMn$_2$O$_4$, distortion in the crystal field is sufficient to open the gap and the on-site repulsion simply amplifies this gap. An arbitrary tetragonal distortion ($c/a = 1.2$) in ZnNi$_2$O$_4$ also opens a gap as can be seen in Fig.~\ref{fig:spinel-distort}. It may be noted that ZnNi$_2$O$_4$ is yet to be synthesized. The distortion is carried out over the hypothetical cubic ZnNi$_2$O$_4$ .

\subsection{Olivine Phosphates: LiMPO$_4$}

\begin{figure}
\centering
\includegraphics[scale=0.37]{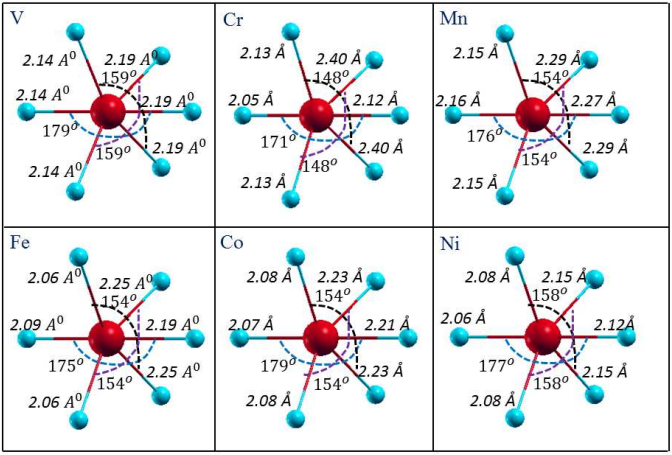}
\caption{The shape of the theoretically optimized MO$_6$ complexes in LiMPO$_4$ (M = V, Cr, Mn, Fe, Co and Ni). Both the bond angles and bond lengths are indicated in the figure. The theoretically optimized structures are in excellent agreement with the experimental one except for the compounds with M = V and Cr which are yet to be synthesized. Transition metal atoms are shown in red.}
\label{fig:olivine-bondlength}
\end{figure}

The monoxides present a perfect octahedra and in the case of Spinels, the octahedra is distorted through the O-M-O bond angles. Also while the monoxides have linear M-O-M chains, the spinels have zig-zag M-O-M chains. In this section we will focus on the family of Olivine phosphates (LiMPO$_4$) which have highly distorted MO$_6$ complexes where both M-O bond lengths and O-M-O bond angles are anisotropic as shown in Fig.~\ref{fig:olivine-bondlength}. Also these family of compounds do not exhibit any M-O-M chains. It has been earlier reported that the lowering in symmetry is driven by the presence of  PO$_4^{3-}$ polyanions \cite{ajit}. Experimental studies show that each member of this family is an antiferromagnetic insulator with conductivity in the range 10$^{-4}$ to 10$^{-11}$ S cm$^{-1}$ \cite{mays, jun, santoro, santoro1, delacourt, baker, allen, babu}. The N\'{e}el temperature is in the range 22 to 53K \cite{jun}. 

\begin{figure*}
\centering
\includegraphics[scale=0.5]{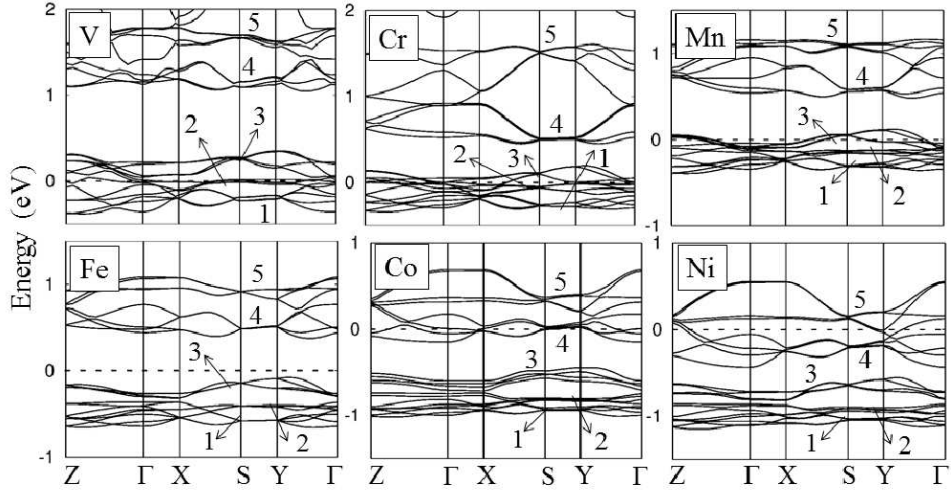}
\caption{Non-magnetic band structure of LiMPO$_4$ (M= V, Cr, Mn, Fe, Co and Ni) using GGA only calculation. E$_F$ is set to zero.}
\label{fig:olivine-nm-bands}
\end{figure*}

There is a minor variation in the crystal structure with change in the transition metal ion. Fig.~\ref{fig:olivine-bondlength} shows that the change in the individual M-O bond length ($\Delta d_i$) across the compounds is in the range 0.01 to 0.25 \AA{} and the change in individual O-M-O bond angle ($\Delta \theta_i$), measured with reference to an O-M-O octahedral axis, lies in the range 0 to 11$^\circ$. We note that similar variations in the bond length and bond angle are observed in the family of pure and doped perovskites AMO$_3$ (A =Sr, Ca, La) \cite{booth, inoue, zhou1, li} which are higher symmetry compounds than LiMPO$_4$.  However, unlike olivines, in pervoskites these minor variations ($\Delta d_i$, $\Delta\theta_i$) with respect to the change in the valence electron concentration (VEC) create a rich electronic and magnetic phase diagram \cite{sashi, booth, inoue, zhou1, george}. For example, if we compare the JT distorted orthorhombic LaMnO$_3$ and LaCrO$_3$, the $\Delta d_i$  and $\Delta\theta_i$  are 0.2 \AA{} and 2$^\circ$ respectively \cite{booth, zhou2, li}. Here, while the Cr based compound is found to be G-type antiferromagnetic, the Mn one stabilizes in type-A antiferromagnetic configuration. Similarly, in the series of SrMO$_3$ (M = V, Cr and Mn) the physical properties vary from metallic to insulator even though $\Delta d_i$ is in the range of 0.02 - 0.04 \AA{} and $\Delta \theta_i$ is zero \cite{inoue, mandal, zhou1, lee1, george} . Therefore, the family of olivine phosphates provides the perfect platform to generalize the electronic and magnetic phases of highly asymmetric 3d transition metal oxides.

To start with we shall discuss the effect of crystal field exerted by O$^{2-}$ ligands on the M-d states by examining the NM GGA band structure of LiMPO$_4$ in the vicinity of E$_F$ plotted in Fig.~\ref{fig:olivine-nm-bands}. The figure suggests that irrespective of M, the bands are narrow which indicates weak covalent hybridization. The figure also infers that in each of the band structure, there is a gap of magnitude 0.3 to 0.8 eV that separates lower lying three set of bands (indicated as 1, 2, and 3) and upper lying two set of bands (indicated as 4 and 5). Further, each set, consisting of four bands due to the four formula unit primitive cell, is well separated from the neighboring sets. The cause of such separation is explained schematically in Fig.~\ref{fig:cf}.  Had the MO$_6$ complex been a perfect octahedral, the five-fold degenerate states would have split into three-fold t$_{2g}$ and two-fold e$_g$ states with separation value $\Delta_{cr}$. However, since there is a significant asymmetry in the MO$_6$ complex (see Fig.~\ref{fig:olivine-bondlength}) there is a new asymmetric crystal field ($\Delta_{cr}'$) that removed both the three-fold and two-fold degeneracy.

\begin{figure*}
\centering
\includegraphics[scale=0.5]{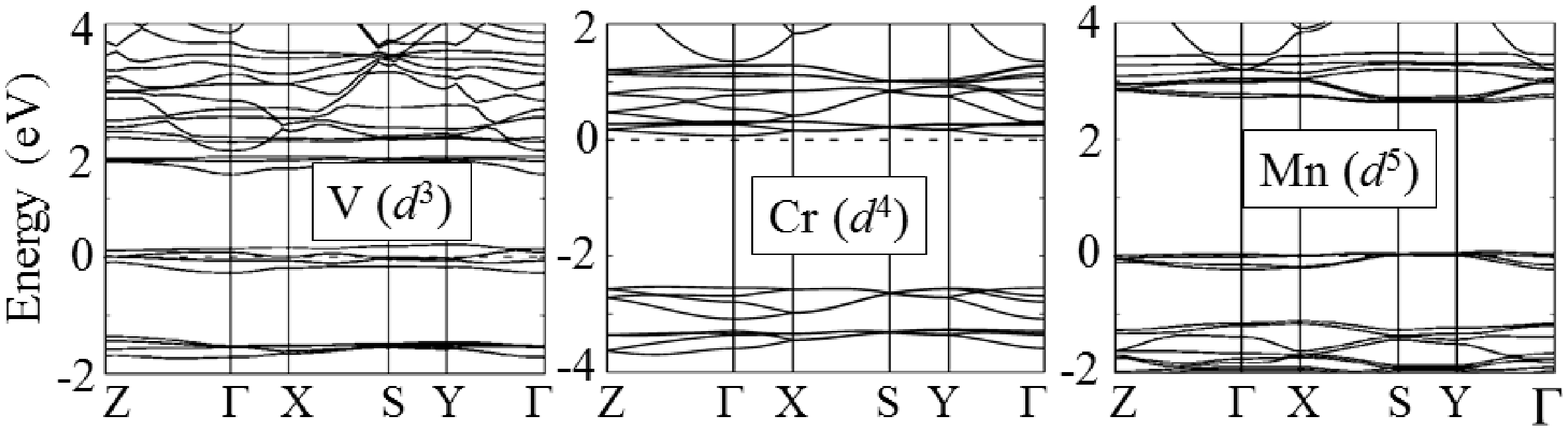}
\caption{Non-magnetic GGA+U (U = 5 eV) band structure of LiMPO$_4$ (M= V, Cr and Mn). For the case of even d electrons, the vanishing gap (within GGA) becomes prominent with inclusion of U.}
\label{fig:olivine-nm+u-bands}
\end{figure*}

In the NM phase (Fig.~\ref{fig:olivine-nm-bands}), odd electron configurations (d$^3$, d$^5$ and d$^7$) show band crossing at E$_F$ due to half-occupied states. On the other hand with even electron configurations (d$^4$, d$^6$ and d$^8$), each state is occupied and therefore either there is a vanishing gap due to small value of $\Delta^{\prime}_{cr}$ or a strong gap as in the case of Fe (d$^6$) where the first three states are occupied and the other two unoccupied states are well separated due to a reasonable $\Delta_{cr}$.  A strong on-site repulsion (U), among the electrons of different orbitals amplifies the vanishing gap for the even electron configuration (see Fig.~\ref{fig:olivine-nm+u-bands}: LiCrPO$_4$). However, the odd electron systems remains gap-less.

\begin{figure*}
\centering
\includegraphics[scale=0.55]{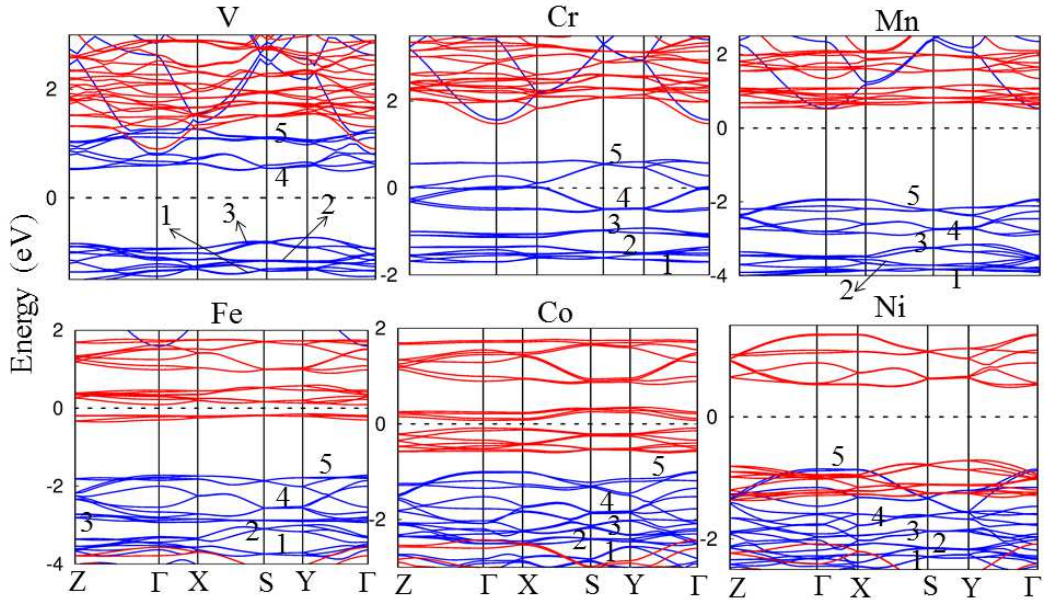}
\caption{Ferromagnetic GGA band structure of LiMPO$_4$. The spin-up and spin-down bands are represented in blue and red respectively. Each of the compounds stabilizes in HS state. Irrespective of M, a band gap mediated by $\Delta_{cr}'$ exists.}
\label{fig:olivine-fm-band}
\end{figure*}

\begin{figure}
    \centering
    \includegraphics[scale =0.5]{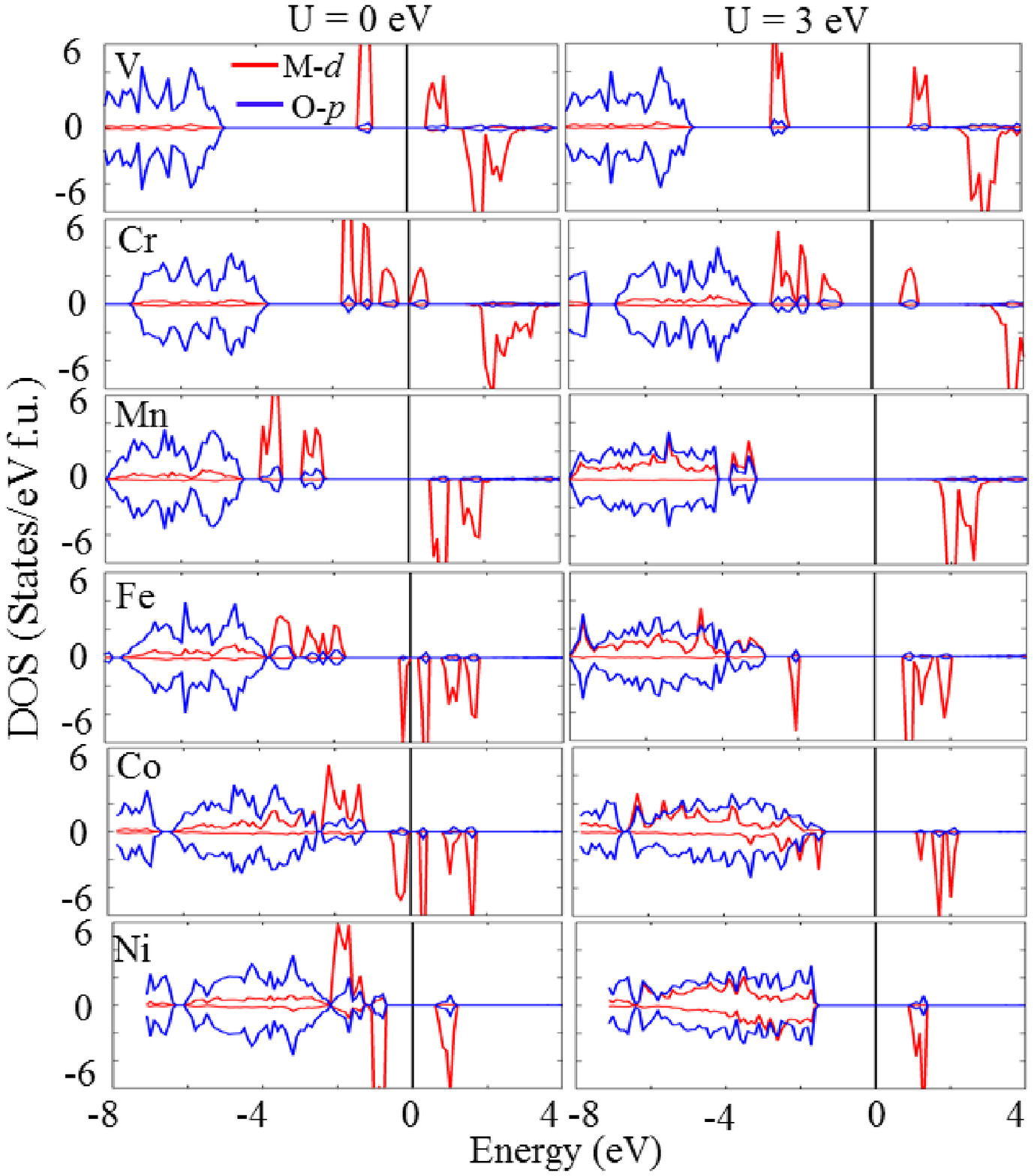}
    \caption{GGA and GGA+U (U = 3 eV) AFM spin resolved DOS for olivine phosphate family.}
    \label{fig:olivine-afm-u-0-3}
\end{figure}

 The spin-polarized band structure, calculated using FM ordering, are shown in Fig.~\ref{fig:olivine-fm-band}. Each member stabilizes in HS state with a reasonable $\Delta_{ex}$ which is always greater than the $\Delta_{cr}'$. Since, unlike to the NM phase, there is no forced half-occupancy due to spin-polarization in the FM phase, $\Delta_{cr}'$ ensures a gap between the occupied and unoccupied states. In general the gap magnitude is either determined by $\Delta_{ex}$ (LiMnPO$_4$, d$^5$), $\Delta_{cr}$ (LiVPO$_4$, d$^3$; LiNiPO$_4$ d$^8$) or $\Delta_{cr}'$ (LiCrPO$_4$, d$^4$; LiFePO$_4$, d$^6$; and LiCoPO$_4$, d$^7$).

If the d states are localized due to negligible p-d and d-d hopping or strong correlation effects, the unpaired spins behave like classical spin to stabilize the AFM ordering. In Fig.~\ref{fig:olivine-afm-u-0-3}, we have shown the GGA and GGA+U calculated M-d DOS in the olivine AFM ordering \cite{ajit}. Like the FM phase, in AFM phase also $\Delta^{\prime}_{cr}$ ensures a minimum gap at E$_F$  between two groups of weakly dispersive bands. Inclusion of on-site repulsion will further enhance the gap as can be seen from Fig.~\ref{fig:olivine-afm-u-0-3}.

\subsection{Summary of the DFT results}

\begin{figure*}
\includegraphics[scale=0.52]{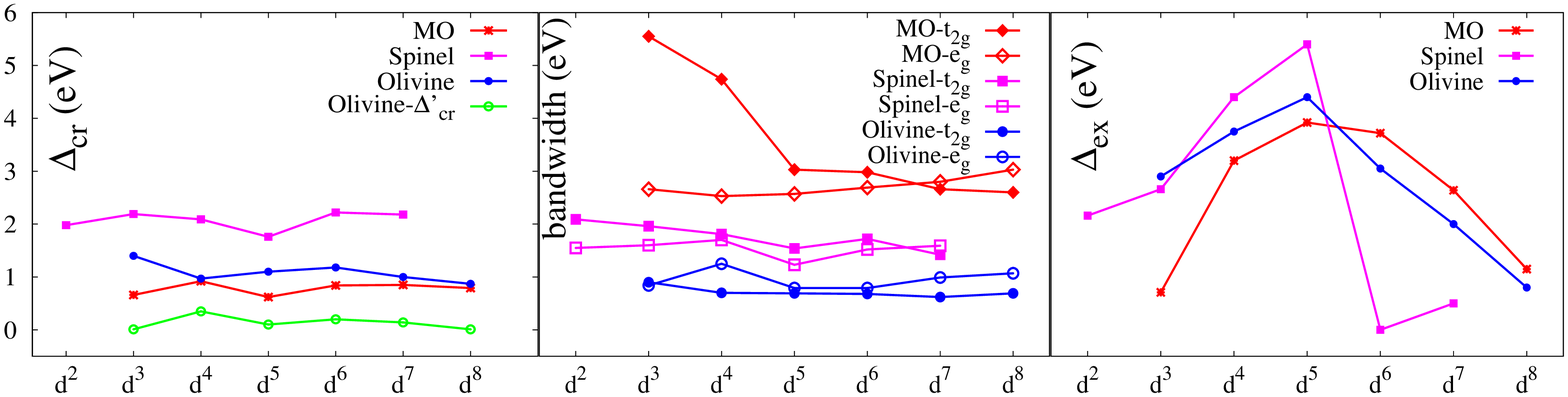}
\caption{Comparison of $\Delta_{cr}$, electron hopping strength measured through band width, and $\Delta_{ex}$ for all the three families of TMOs discussed in this work. The results are obtained with U = 0. For the olivine family $\Delta{^\prime}_{cr}$ is also shown.}
\label{fig:splitting}
\end{figure*}

To develop an universal phenomenology over the relation between ground state electronic structure and the symmetry dependent coupling parameters of the MO$_6$ complex, in Fig.~\ref{fig:splitting} we have shown the variation of $\Delta_{cr}$, t$_{2g}$ and e$_g$ band width, and $\Delta_{ex}$ as a function of d$^n$ across the three TMO families. The $\Delta_{cr}$ is found to be about 0.8 - 1 eV higher in spinel members than the monoxide and olivine members. It may be noted that the average M-O bond length of the MO$_6$ complex in the spinel family is nearly 0.2 \AA{} smaller than that of the monoxide and olivine family. As expected the complete breakdown of the octahedral symmetry in  olivine compounds makes all the d-states non-degenerate through $\Delta{^\prime}_{cr}$ which is approximately 0.2 eV. With reduction in the symmetry of the M-O complex, the p-d covalent interaction diminishes and therefore, t$_{2g}$ and e$_g$ band widths decreases and hence, monoxides exhibit maximum bandwidth and on the other hand the asymmetric olivines creates atomically localized states. The spin-split $\Delta_{ex}$ follows the Hund's rule and therefore, it is maximum for  half-filled shell and decreases with further increase or decrease in the electron occupancy. However, deviation is observed in the case of ZnCo$_2$O$_4$ and ZnNi$_2$O$_4$, where large $\Delta_{cr}$ and reduced band width ensures LS spin configuration of these compounds in the independent electron approximation. 

Further our analysis suggests the following two important trends: (I) When a TMO exhibit very narrow band width at E$_F$ in the independent electron approximation (e.g. LiMPO$_4$), it tends to become insulating irrespective of the magnetic ordering. In these compounds, at low temperature the unpaired spins behave like classical spins to order antiferromagnetically. (II) Strong correlation effect modifies the electronic structure substantially if in the independent electron approximation, a d sub-band (e.g. t$_{2g}$, e$_g$) with large band width partially occupies the E$_F$. Such systems are in the intermediate stable phase become correlated metal (CM) (e.g., ZnV$_2$O$_4$, ZnNi$_2$O$_4$ and CrO). Under appropriate conditions they may create well separated lower and upper Hubbard bands to make the system a correlated insulator (CI) (e.g., FeO, CoO and ZnCo$_2$O$_4$). Also such compounds are prone to structural transition at lower temperature. Either they undergo JT mediated minor lattice distortion (e.g. CrO, ZnV$_2$O$_4$, see Fig.~\ref{fig:spinel-distort})  or large scale structural transition (ST) to remove the degeneracy at E$_F$ which in turn opens a gap. In the literature, ST is observed in ZnMn$_2$O$_4$ (cubic to tetragonal) \cite{choi} to break the e$_g\uparrow$ degeneracy and in CuO (cubic to distorted square-planar) \cite{ganga} to remove the t$_{2g}\downarrow$ degeneracy.

Literature  suggests that if a TMO consists of multiple transition metal ions, then the net electronic structure is primarily a sum of electronic structure of individual metal-oxygen complexes. For example, in the case of NiFe$_2$O$_4$, there are three complexes: NiO$_6$, FeO$_6$ and FeO$_4$. The electronic structure calculations \cite{ugendar} show that each metal-oxygen complex retains its individual characteristics, except that the FeO$_4$ local spin moment is antiferromagnetically coupled to the NiO$_6$ and  FeO$_6$ local spin moments. Similar is the case of ferromagnetic insulator LaCoMnO$_6$ \cite{baidya},where MnO$_6$ octahedra has a d$^3$ configuration and its electronic structure resembles to that of VO [Fig.~\ref{fig:tmmo-afm}] and CoO$_6$ octahedra has a d$^7$ configuration and its electronic structure resembles to that of CoO (Fig.~\ref{fig:tmmo-afm}). Therefore, if the electronic configuration of individual M-O complexes are known, a firsthand information on the electronic properties of a given TMO can be obtained.

\section{Empirical Hypotheses}
\label{sec:schematics}
The understanding that we have developed from analyzing the spin-polarized band structure of ideal octahedra and pseudo-octahedra as well as asymmetric MO$_6$ complexes can further be extrapolated to any generic M-O complexes to derive an universal phenomenology, in the shape of a periodic chart, which can be applied to any 3d TMOs in explaining their electronic and magnetic structure. The phenomenology is defined through the competition between $U$, $\Delta_{cr}$ ($\Delta^{\prime}_{cr})$, $\Delta_{ex}$, and electron filling d$^n$.

\subsection{Electronic and magnetic structure of octahedral 3d TMOs}
\begin{figure*}
\centering
\hspace*{-0.4cm}\includegraphics[scale=0.425]{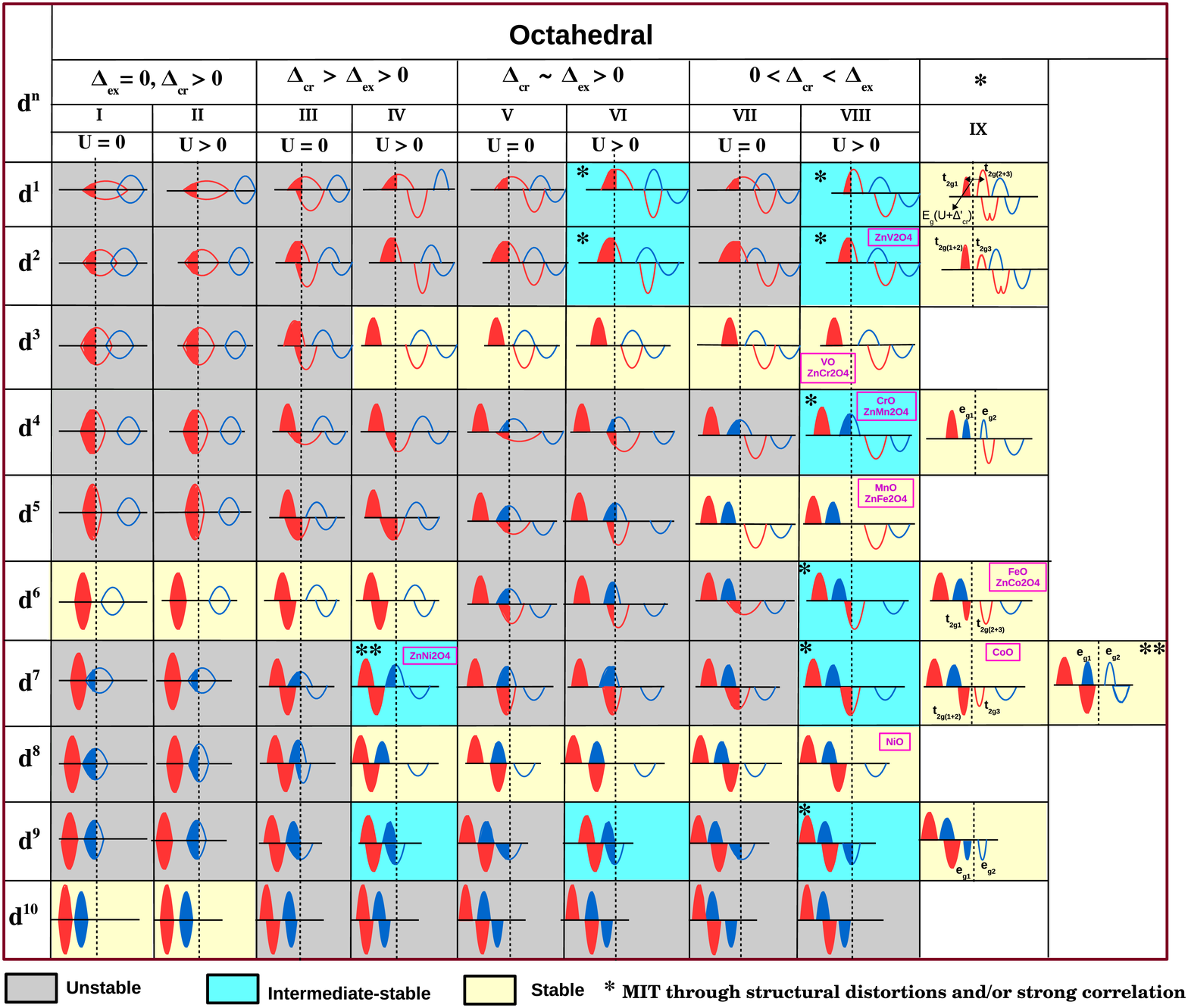}
\caption{Schematic DOS representing the electronic and magnetic structure of octahedral TMOs as a function of electron filling (d$^n$), crystal field split ($\Delta_{cr}$), spin-split ($\Delta_{ex}$), and onsite Coulomb repulsion (U). The $t_{2g}$ and $e_g$ DOS are colored as red and blue respectively. While the value of the aforementioned parameters are compound specific, only comparative domains are shown to demonstrate generalized features. For example, U $>$ 0 indicates a reasonable value which properly represents the correlation effect and can vary from compound to compound. The columns III and IV represent the scenario in which $\Delta_{ex}$ is substantially lower than $\Delta_{cr}$ and reverse is the scenario for columns VII and VIII. The electronic contribution by the M-O complex towards stability of the crystal is represented through the background color of the concerned box. The intermediate stable states (cyan background) represents correlated metals (CM)  and can make a transition to insulating stable phase as shown in column IX. The transition can occur through minor Jahn-Teller (JT) distortion which is highly probable or through large-scale structural transition (ST) which is less probable. The ground state electronic structures of the octahedral compounds examined in this paper are indicated.  The Metal insulator transition (MIT) can also be achieved only through strong correlation as observed in the case of FeO and ZnCo$_2$O$_4$. Such compounds are classified as correlated insulators (CIs).}
\label{fig:octahedral}
\end{figure*}

Fig.~\ref{fig:octahedral} schematically presents the electronic structure of MO$_6$ octahedra through M-d DOS as a function of the aforementioned parameters. For d$^1$ systems, in the absence of spin-exchange split and on-site repulsion, the partially occupied t$_{2g}$ states will behave like itinerant electrons to yield a non-magnetic metallic state as shown in column I. With a substantial $\Delta_{cr}$, increase in $\Delta_{ex}$ and U, the localization increases as shown in columns VI and VIII to make the system a correlated metal (CM) which is observed for Ca$_{1-x}$Sr$_x$VO$_3$ \cite{inoue}. However, high DOS at E$_F$ makes the 3d$^1$ systems prone to metal insulator transition (MIT) through minor JT based structural transition (ST) as demonstrated in column IX (first row) and observed in many systems such as LaTiO$_3$ \cite{Dymkowski, arima} and SrVO$_3$ \cite{morikawa}. The NM metallic behavior is observed in higher temperature once the partially occupied state overcomes the localization barrier.

As the second row of Fig.~\ref{fig:octahedral} suggests, the d$^2$ systems will behave like the d$^1$ systems. The observed d$^2$ JT-ST mediated insulator is LaVO$_3$ \cite{arima}. The d$^3$ systems makes the t$_{2g}$ states half-occupied and therefore according to Hund's coupling, the systems will stabilize with large $\Delta_{ex}$. Even in the absence of strong correlation effect, they will exhibit insulating behavior as can be seen from columns V and VII of the third row. The DOS of ZnCr$_2$O$_4$ shown earlier in Fig.~\ref{fig:spinel-afm} is testimony to it. Such structures have are generally stable with regular octahedra. However, the regular octahedra may tilt to bring a distortion to the crystal structure as in the case of CaMnO$_3$ \cite{zhou3} where the regular octahedra are tilted with respect to the neighboring octahedra to make the structure orthorhombic. The strength of the tilting of the octahedra varies with applied pressure and temperature. 

Like the d$^3$ systems, d$^4$ systems prefer to stabilize in HS configuration (t$^{3}_{2g}\uparrow$ e$^{1}_{g}\uparrow$) with large $\Delta_{ex}$. Such systems generally undergoes the JT transitions to split the double degenerate e$_g$ states, e.g., CrO, (Fig.~\ref{fig:cro-jt}), and ZnMn$_2$O$_4$, (Fig.~\ref{fig:spinel-distort}) and LaMnO$_3$ \cite{sashi} form a correlated insulator (CI) as shown in column IX of fourth row. If the Q$_3$ mode of distortion prevails, the cubic to tetragonal transition occurs (e.g. ZnMn$_2$O$_4$ \cite{choi}) and if both Q$_2$ and Q$_3$ modes of distortion prevails, a cubic to orthorhombic transition is expected \cite{nanda, nanda1}. Maximum $\Delta_{ex}$ is observed in d$^5$ systems, where both t$_{2g}$ and e$_g$ states are half-occupied. These systems are primarily antiferromagnetic insulators (MnO (Fig.~\ref{fig:tmmo-afm}), ZnFe$_2$O$_4$ (Fig.~\ref{fig:spinel-afm})), and exhibit maximum stability. In few compounds like quasi-one dimensional spin-chain oxides Ca$_3$ZnCoO$_6$ \cite{jayeeta}, where $\Delta_{cr} > \Delta_{ex}$ and correlation effect is weak, the LS state (Co - t$^{3}_{2g}\uparrow$ t$^{2}_{2g}\downarrow$) stabilizes to make the compound a correlated metal.

The d$^6$ and d$^7$ systems can stabilize in LS configuration to become non-magnetic and weakly magnetic respectively (ZnCo$_2$O$_4$ and  ZnNi$_2$O$_4$ (Fig.~\ref{fig:spinel-afm})). However, systems with strong correlation effect and weak crystal field can be stabilized in the HS state as in case of FeO, CoO, ZnCo$_2$O$_4$ which is discussed earlier. The LS d$^7$ systems has half-occupied e$_g$ states in the spin-majority channel suggesting possible JT distortion. In fact, in Fig.~\ref{fig:spinel-distort}, we have shown that such a distortion makes ZnNi$_2$O$_4$ insulating. In the HS state, the spin-down d$^6$ and d$^7$ electronic structures replicate the spin-up electronic structures of d$^1$ and d$^2$ systems. It may be noted that the gap can be created without structural distortion through strong correlation as seen in the case of FeO and CoO (Fig.~\ref{fig:tmmo-afm}). Such systems are ideal correlated insulators and rare. The d$^8$ systems, e.g. NiO, has the stable electronic configuration of t$^{3}_{2g}\uparrow$ t$^{3}_{2g}\downarrow$ e$^{2}_{g}\uparrow$ and like d$^3$ systems, it favours perfect octahedra. The spin-down d$^9$ electronic structure replicates the spin-up electronic structures of HS d$^4$ configuration. The d$^{10}$ systems are non-magnetic insulators by virtue of their filled d-shells.

\subsection{Electronic and magnetic structure of tetrahedral 3d TMOs}

\begin{figure*}
\hspace*{-0.5cm}\includegraphics[scale=0.42]{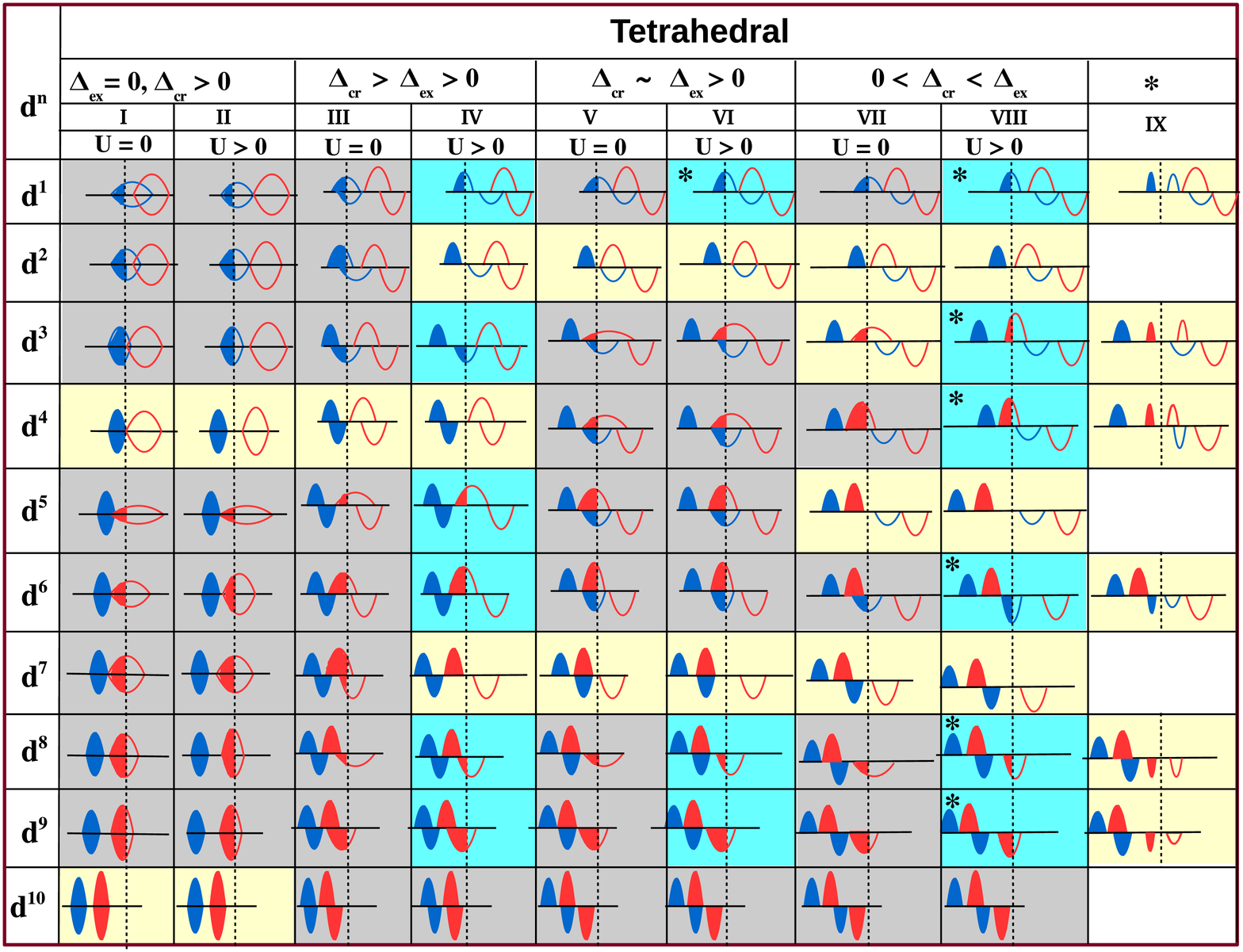}
\caption{Schematic DOS representing the electronic and magnetic structure of tetrahedral TMOs as a function of  d$^n$, $\Delta_{cr}$, $\Delta_{ex}$, and U. For detail description see caption of Fig.~\ref{fig:octahedral}.}
\label{fig:tetrahedral}
\end{figure*}

The tetrahedral complex with T$_d$ point group symmetry exerts a $\Delta_{cr}$ which is roughly 45\% of that of an octahedra with same M-O bond length. In addition, the d-states are now split into $e + t_2$ with the former lying lower in energy. Therefore, while the overall trend of the tetrahedral electronic structure, schematically illustrated in Fig.~\ref{fig:tetrahedral}, remains similar to that of an octahedral, there are subtle differences in creating JT and ST mediated MIT. The lattice mode of distortions are no longer cubic, i.e., in-plane Q$_2$ and axial Q$_3$ modes. Rather they are three dimensional Q$_{\epsilon}$ and Q$_{\epsilon^\prime}$ modes which govern the distortion \cite{bersuker_2006}. The unstable electronic configurations, highlighted in grey background, are self explanatory and need not be discussed. The d$^1$ systems either can be CM or JT-ST mediated CI as observed in YCrO$_4$ \cite{tsirlin}, Ba$_2$Cr$_3$O$_8$ \cite{kofu} and Sr$_2$Cr$_3$O$_8$ \cite{radtke}. The d$_2$ systems are stable HS magnetic insulators. 
The HS d$^3$ and d$^4$ systems behave like octahedral d$^1$ and d$^2$ systems. However, like the d$^6$ octahedral systems, the d$^4$ system can remain stable in LS (e$^2\uparrow$ e$^{2}\downarrow$). The d$^5$ system can be JT active in LS state, provided there is a reasonable $\Delta_{cr}$ else it remains HS magnetic insulator as in case of NiFe$_2$O$_4$ \cite{ugendar}. Irrespective of LS or HS configuration, the tetrahedral d$^6$ system exhibits high DOS at E$_F$ and hence prone to structural distortion to create MIT. The d$^7$ system highly likely to possess perfect tetrahedra with e$^{2}\uparrow$ e$^{2}\downarrow$ t$^{3}_{2}\uparrow$ electronic configuration with large band gap, e.g., Co$_3$O$_4$ \cite{chen1}. The d$^8$ and d$^9$ systems have partially occupied t$_2$ states in the spin minority channel. Therefore they are either CMs (in the intermediate-stable state) or  CIs mediated through strong on-site repulsion with or without JT distortion. 

\begin{figure*}
\hspace*{-0.5cm}\includegraphics[scale=0.4]{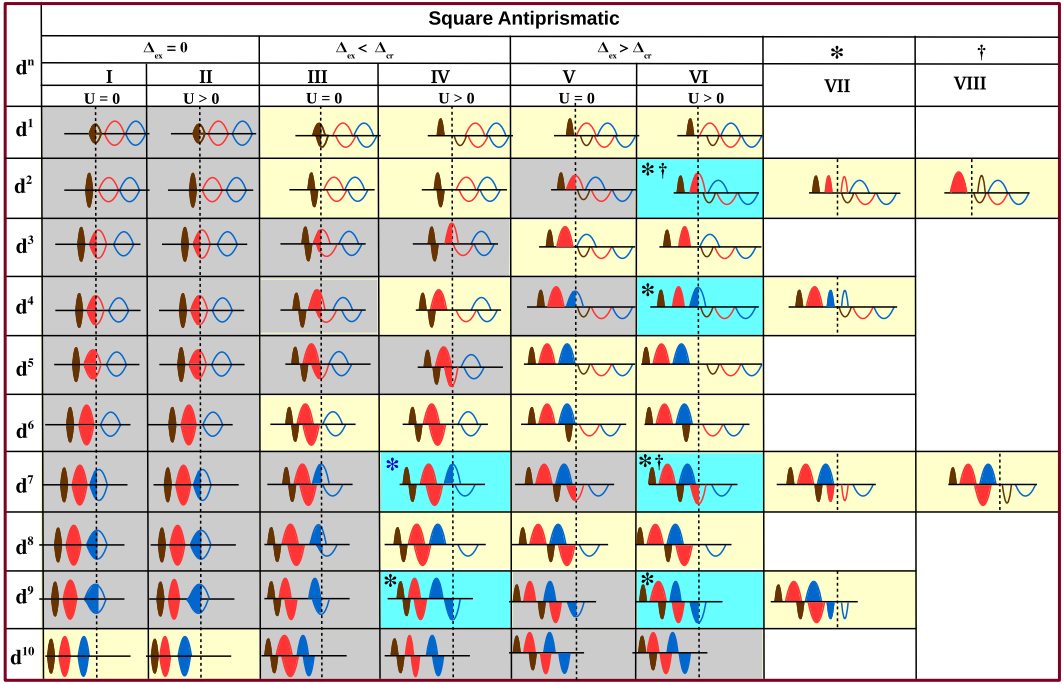}
\caption{Schematic DOS representing the electronic and magnetic structure of square antiprismatic TMOs as a function of  d$^n$, $\Delta_{cr}$, $\Delta_{ex}$, and U. The DOS in brown, red and blue colors represent $a_1$ ($d_{z^2}$), $e_2$ ($d_{xy}$, $d_{x^2-y^2}$), and $e_3$ ($d_{yz}, d_{xz}$) states respectively. For detail description see caption of Fig.~\ref{fig:octahedral}.  The d$^2$ and d$^7$ systems become insulating either through lattice distortion (indicated by $\ast$) or through swapping of the a$_1$ and e$_2$ order (indicated by $\dagger$). The latter will occur when increase in elastic energy is more than the $\Delta^{\prime}_{cr}$ between a$_1$ and e$_{2}$. The d$^7$ LS configuration (column-IV, indicated by blue star) can also undergo a MIT transition through lattice distortion similar to that of HS configuration. }
\label{fig:sq_anti}
\end{figure*}

\subsection{Electronic and magnetic structure of square antiprismatic 3d TMOs}

The square antiprismatic complex with D$_{4d}$ symmetry splits the d states into three levels. In the increasing energy order they are a$_1$ (d$_{z^2}$), two fold degenerate e$_2$ (d$_{x^2-y^2}$ and d$_{xy}$), and two fold degenerate e$_3$ (d$_{xz}$ and d$_{yz}$). In terms of octahedral field split ($\Delta^{oct}_{cr}$), the strength of split between a$_1$ and e$_2$ is  approximately 0.4$\Delta^{oct}_{cr}$ and that between e$_2$ and e$_3$ is approximately 0.45$\Delta^{oct}_{cr}$. The electronic structure of such a complex is schematically illustrated in Fig.~\ref{fig:sq_anti}. Most of them are self explanatory. Following are the two most interesting features exhibited by the square-antiprismatic complex.  (I) The d$^2$, d$^4$ and d$^6$ can stabilize both in LS and HS configuration. However, the HS configuration may lead to J-T type structural distortion to weaken correlated metallic behavior. (II) The HS d$^2$ and d$^7$ systems can also become insulators through swapping of the a$_1$ and e$_2$ energy order. Such a situation will occur when for a given compound, the elastic energy due to distortion is far more than the crystal field split between a$_1$ and e$_2$ states.  It may be noted that there is hardly any literature which reports TMOs with square antiprismatic complexes. However, if such TMOs are synthesized, they are highly likely to exhibit the properties as suggested in Fig.~\ref{fig:sq_anti}.

Like the square antiprismatic MO complexes, D$_{3h}$ symmetry based trigonal prismatic and trigonal bipyramidal MO complexes also splits the five-fold degenerate d-states into three levels. For these complexes, schematic electronic structure diagram similar to Fig. 20 can be drawn. They are not shown here to avoid the redundancy.

\subsection{Electronic and magnetic structure of square-planar and completely asymmetric 3d TMOs}

\begin{figure*}
\centering
\includegraphics[scale=0.4]{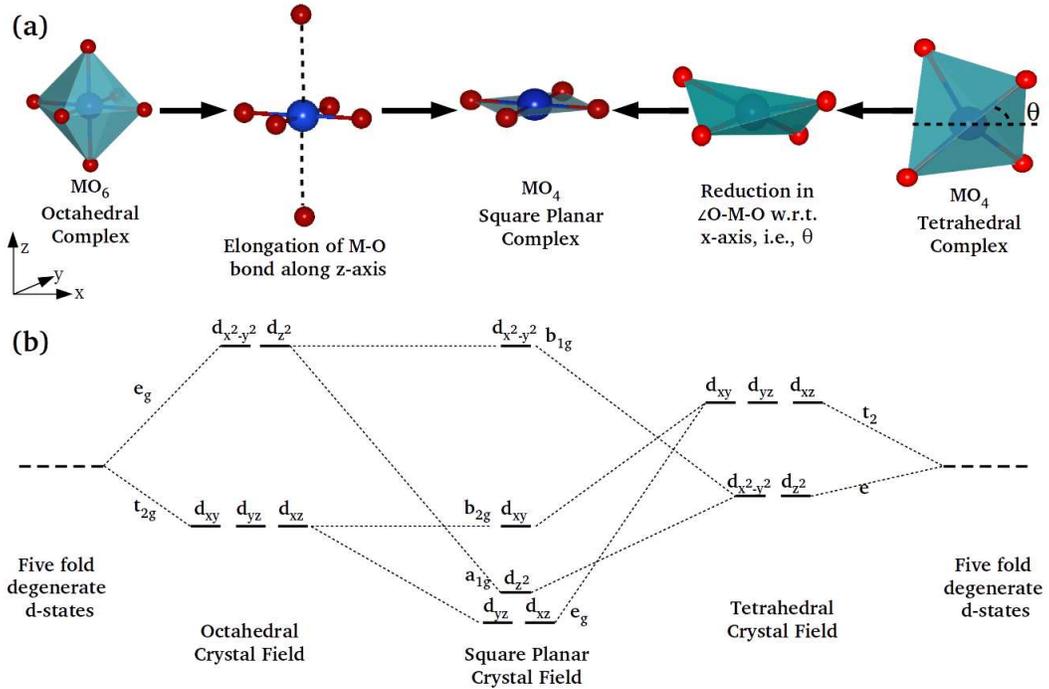}
\caption{(a) Symmetry breaking from octahedral to square planar (left to center) and tetrahedral to square planar structure (right to center). (b) Crystal field splitting from octahedra to square planar and tetrahedra to square planar.}
\label{fig:sym-break}
\end{figure*}

\begin{figure*}
\includegraphics[scale=0.4]{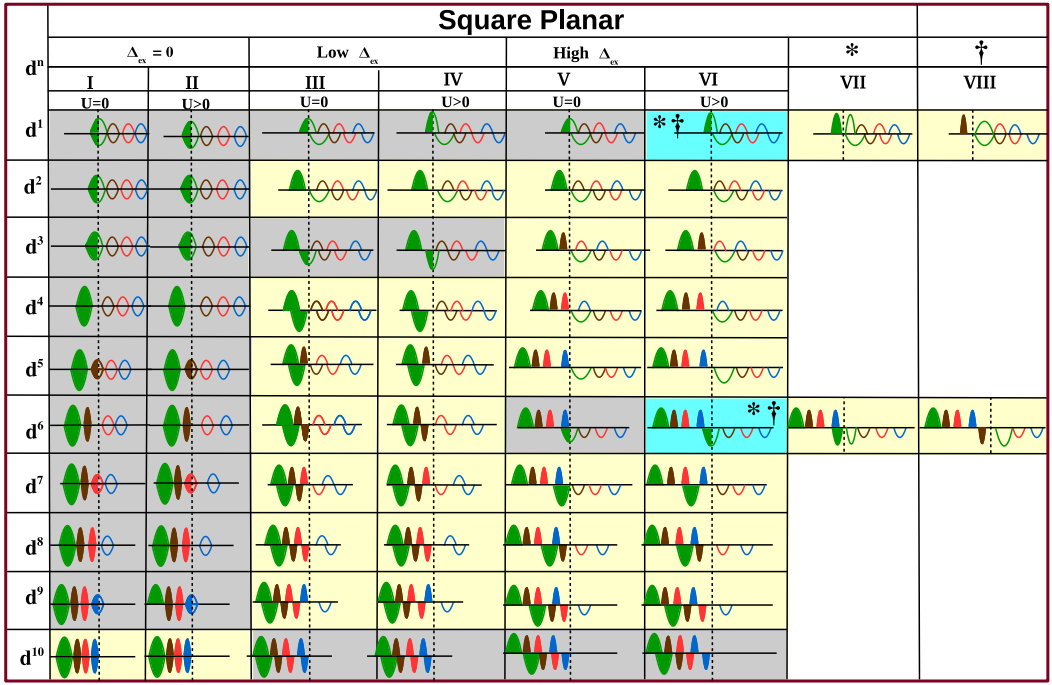}
\caption{Schematic DOS representing the electronic and magnetic structure of square-planar TMOs as a function of d$^n$, $\Delta_{cr}$, $\Delta_{ex}$, and U. Here green, brown, red and blue colors represent e$_g$, a$_{1g}$, b$_{2g}$ and b$_{1g}$ DOS, respectively. For detail description see caption of Fig.~\ref{fig:octahedral}. The d$^1$ and d$^6$ systems become insulating either through lattice distortion (indicated by $\ast$) or through swapping of the e$_g$ and a$_{1g}$ order (indicated by $\dagger$). The latter will occur when increase in elastic energy is more than the $\Delta^{\prime}_{cr}$ between e$_g$ and a$_{1g}$ }
\label{fig:sq_pl}
\end{figure*}

\begin{figure*}
\centering
\hspace*{-0.3cm}\includegraphics[scale=0.34]{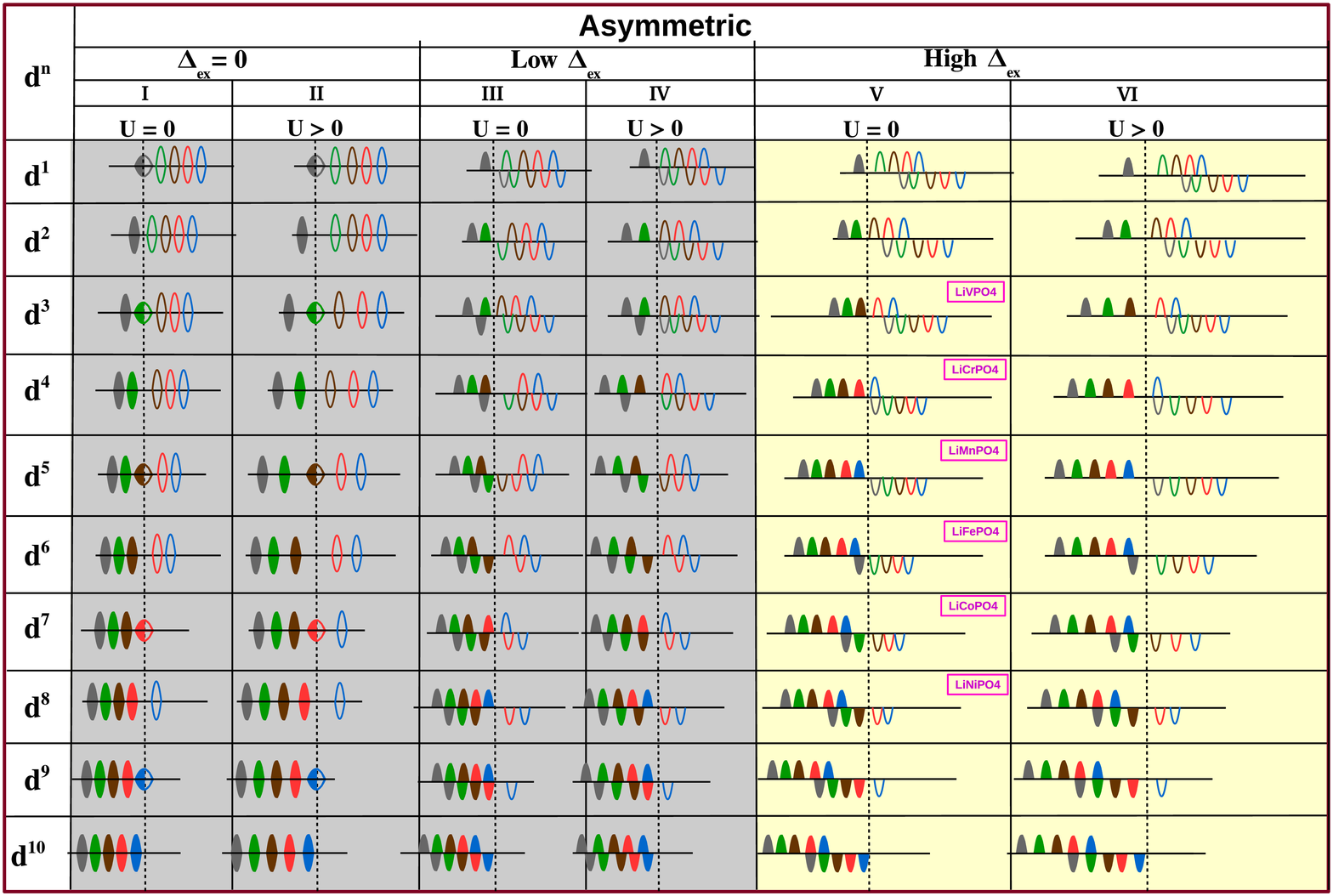}
\caption{Schematic DOS representing the electronic and magnetic structure of asymmetric TMOs as a function of d$^n$,  $\Delta^{\prime}_{cr}$,  $\Delta_{ex}$, and U. Five non-degenerate d states are represented in five different colors. For detail description see caption of Fig.~\ref{fig:octahedral}.}
\label{fig:five-fold}
\end{figure*}

The discussion on the schematic demonstration of the electronic structure of octahedral and tetrahedral M-O complexes field symmetry is further extended to  square-planar M-O complexes. In fact the latter is realized through symmetry breaking as can be seen from Fig.~\ref{fig:sym-break} (a). This symmetry breaking essentially further split 3-fold and 2-fold degenerates into one two-fold degenerate and remaining three non-degenerate states as shown in Fig.~\ref{fig:sym-break} (b). The reduced symmetry weakens the p-d and d-d covalent interactions to make the states narrower. With or without correlation and with reasonable $\Delta_{ex}$ band gap at E$_F$ will appear in most of the cases except in d$^1$, d$^3$ and d$^6$ systems (Fig.~\ref{fig:sq_pl}). The LS d$^1$ and HS d$^6$ show half-occupied e$_g$ states in the spin-majority and minority channels respectively. Under this condition, two kinds of transitions may occur. Either the structure will distort to create MIT (see column VII) or through state swapping (see column VIII) where the a$_{1g}$ (d$_z^2$)  swap with the e$_g$ (d$_{xz}$ + d$_{yz}$) states. It may be noted that in the high-temperature non-magnetic phase the a$_{1g}$ state will remain partially occupied with itinerant electrons for the d$^5$ system to enhance the conductivity along $z$-axis. Similarly the conductivity can be large along the $xy$-plane for the d$^9$ system. However, at low temperature, these structure will be unstable and further distortion can occur which has been observed earlier in CuO by one of us \cite{ganga}.

Further break down of symmetry basically makes all the five d-states non-degenerate as is observed in the family of olivine phosphates which is discussed in detail in section~\ref{sec:DFT-ES} (C). These systems create extremely localized states resembling to that of molecules and clusters as schematically shown in Fig.~\ref{fig:five-fold}. In the high temperature non-magnetic phase, odd electron systems (d$^1$, d$^3$, d$^5$,d$^7$ and d$^9$) will show correlated metallic behavior with forced half-occupancy (see Fig.~\ref{fig:five-fold}). With lowering in the temperature, a minor $\Delta_{ex}$ makes them magnetic insulating. The even electron systems are always insulators. The strength of magnetization depends on the strength of $\Delta_{ex}$ and electron filling of the the d-shell. In the low temperature generally the unpaired spins are either weakly antiferromagnetically coupled or have random orientation to stabilize in the paramagnetic phase.

\section{Summary and Conclusion}
\label{sec:conclusion}
 
With the intent of developing an universal phenomenology governing the electronic structure of 3d transition metal oxides, the first principles studies were carried out on three families of compounds: monoxides (MO), spinels (ZnM$_2$O$_4$) and olivine phosphates (LiMPO$_4$), where M is a 3d transition metal element. Each of these compounds have MO$_6$ complexes. However, in monoxide it is perfect octahedra to produce the point group symmetry of highest order. In spinels, the MO$_6$ complexes create pseudo-octahedra and in olivine phosphates they are completely asymmetric. Therefore, analysis of the electronic structure of these three families guided us in understanding the effect of the symmetry of the M-O complex on the electronic structure.  In addition,  varying M from V to Ni helped in understanding the effect of d-band filling on the transport properties of these oxides. We find that the ground state is determined from the collective influence of on-site Coulomb repulsion (U), crystal symmetry dependent parameters: electron hopping integral (t) and crystal-field split ($\Delta_{cr}$), and electron-filling dependent spin-exchange split ($\Delta_{ex}$).

The understanding over the octahedral compounds are extrapolated with substantial support from the literature data to develop empirical hypotheses. These hypotheses, shown through schematic DOS as a function of the aforementioned parameters in Figs.~\ref{fig:octahedral},~\ref{fig:tetrahedral},~\ref{fig:sq_anti},~\ref{fig:sq_pl} and~\ref{fig:five-fold} can be used to generate robust firsthand information on the spin-resolved electronic structure of 3d transition metal oxides. For example our hypotheses can successfully predict the conditionalities that produce correlated metals, correlated insulators, JT mediated metal-insulator transitions as well as large scale structural transitions. This work provides the direction to construct similar hypothesis for 4d transition metal oxides and 4f rare earth compounds. In the latter case, it is necessary to include the effect of spin-orbit coupling.

It is necessary to note that such an exploration is unique and has not been attempted before. However, by no means it is complete and therefore opens-up plethora of possibilities. For example, the present study is not sufficient enough to predict the magnetic ordering even though it is accurate enough to explain the local spin moment formation. Therefore, if relation between the spin-spin coupling and arrangement of M-O complexes, which is space-group dependent, can be generalized, the exact electronic and magnetic ground state can be predicted. Also, in the present study the examination of structural stability is confined to the electronic contributions. The role of constituents other than M and O on the structural stability needs to be studied individually. As said in the introduction, the present work focuses only on undoped antiferromagnetic insulating oxides. The doped TMOs often have fractional charge states (i.e. fractional d-orbital occupancy) which can lead to charge ordered and orbital ordered insulators or itinerant double exchange mechanism mediated ferromagnetic metals depending on the dopant concentration. One of the widely studied compound in this context is La$_{1-x}$Sr$_{x}$MnO$_3$ \cite{piskunov, pavone, paraskevopoulos, fang, chen3}. Therefore, in future we shall explore the universality in the electronic structure of TMOs with fractional d-occupancies.

\section*{Acknowledgements}
We would like to acknowledge HPCE computational facility of Indian Institute of Technology Madras. This work is supported by Department of Science and Technology, India through Grant No. EMR/2016/003791. BRKN would like to thank Shantanu Mukherjee for useful discussions.

\bibliography{paper} 

\providecommand{\noopsort}[1]{}\providecommand{\singleletter}[1]{#1}%

\end{document}